\documentclass{aims}
\usepackage{amsmath,bbm}
  \usepackage{paralist}
  \usepackage{graphics} %% add this and next lines if pictures should be in esp format
  \usepackage{epsfig} %For pictures: screened artwork should be set up with an 85 or 100 line screen
\usepackage{graphicx}  
\usepackage{epstopdf}%This is to transfer .eps figure to .pdf figure; please compile your paper using PDFLeTex or PDFTeXify.
\usepackage{subfigure}
 \usepackage[colorlinks=true]{hyperref}
\hypersetup{urlcolor=blue, citecolor=red}

\def\currentvolume{X}
 \def\currentissue{X}
  \def\currentyear{200X}
   \def\currentmonth{XX}
    \def\ppages{X--XX}
     \def\DOI{10.3934/xx.xx.xx.xx}

\newtheorem{proposition}{Proposition}

\theoremstyle{definition}

\renewcommand{\S}{\mathbb{S}}

\graphicspath{{./Figures/}}

\renewcommand{\paragraph}[1]{\bigskip {\bf #1}.}

\date{}

\def\R{\mathbb{R}}

\def \E {\mathbb{E}}

\newcommand{\eps}{\varepsilon}

\def\s{\mathbf{s}}

\newcommand{\f}{\varphi}

\title[The hipster effect] %Use the shortened version of the full title
      {The hipster effect: when anti-conformists all look the same}

\author[Jonathan D. Touboul]{}

 % Oscillation Theory

 \email{jtouboul@brandeis.edu}
\thanks{$^*$ Corresponding author: jtouboul@brandeis.edu}

\begin{document}
\maketitle

% Enter the first author's name and address:
\centerline{\scshape Jonathan D. Touboul$^*$}
\medskip
{\footnotesize
% please put the address of the first author
 \centerline{Department of Mathematics and Volen National Center for Complex Systems, Brandeis University}
   \centerline{415 South Street}
   \centerline{ Waltham, MA 02453, USA}
} % Do not forget to end the {\footnotesize by the sign }

\medskip
\bigskip

% The name of the associate editor will be entered by an editorial staff
% "Communicated by the associate editor name" is not needed for special issue.
 \centerline{(Communicated by the associate editor name)}

%The abstract of your paper
\begin{abstract}

In such different domains as statistical physics, neurosciences, spin glasses, social science, economics and finance, large ensemble of interacting individuals evolving following (mainstream) or against (hipsters) the majority are ubiquitous. Moreover, in a variety of applications, interactions between agents occur after specific delays that depends on the time needed to transport, transmit or take into account information. This paper focuses on the role of opposition to majority and delays in the emerging dynamics in a population composed of mainstream and anti-conformist individuals. To this purpose, we introduce a class of simple statistical system of interacting agents taking into account (i) the presence of mainstream and anti-conformist individuals and (ii) delays, possibly heterogeneous, in the transmission of information. In this simple model, each agent can be in one of two states, and can change state in continuous time with a rate depending on the state of others in the past. We express the thermodynamic limit of these systems as the number of agents diverge, and investigate the solutions of the limit equation, with a particular focus on synchronized oscillations induced by delayed interactions. We show that when hipsters are too slow in detecting the trends, they will consistently make the same choice, and realizing this too late, they will switch, all together to another state where they remain alike. {Another modality synchronizing hipsters are asymmetric interactions, particularly when the cross-interaction between hipsters and mainstreams aree prominent, i.e. when hipsters radically oppose to mainstream and mainstreams wish to follow the majority, even when led by hipsters. We demonstrate this phenomenon analytically using bifurcation theory and reduction to normal form. We find that, in the case of asymmetric interactions, the level of randomness in the decisions themselves also leads to synchronization of the hipsters.}  Beyond the choice of the best suit to wear this winter, this study may have important implications in understanding synchronization of nerve cells, investment strategies in finance, or emergent dynamics in social science, domains in which delays of communication and the geometry of information accessibility are prominent.
\end{abstract}

\section*{Introduction}
\emph{Hipsters avoid labels and being labeled. However, they all dress the same and act the same and conform in their non-conformity. Doesn't the fact that there is a hipster look go against all hipster beliefs?} This perspicacious observation of the blogger Julia Plevin~\cite{plevin:08} ten years ago seems to stand the test of time. Uncovering the structures behind this apparent paradox goes beyond finding the best suit to wear this winter. They can have implications in deciphering collective phenomena in economics and finance, where individuals may find an interest in taking positions in opposition to the majority (for instance, selling stocks when others want to buy), but also, more abstractly, in neuronal networks where high levels of activation inhibitory neurons results in silencing other cells, thereby enforcing opposite reactions on others. 

The question of collective behaviors in large systems of interacting agents taking decisions under uncertainty and based on partial observations belongs to the wide literature of statistical physics. In this domain, models were developed in such different domains as the alignment of spins in magnets~\cite{sherrington-kirkpatrick:75,dai2013curie}, transmission of electrical information in networks of neurons~\cite{crisanti-sompolinsky:87,hermann2012heterogeneous}, and choices in economics and social science~\cite{challet2000modeling,galam2012sociophysics}. In this manuscript, we build upon the wide literature on mean-field systems of interacting agents with finite state spaces, the most classical being the so-called Curie-Weiss model of binary units or the P-state Potts model. Here, we shall shall consider a general model of a population inspired by the Curie-Weiss model, with two twists: 
\begin{itemize}
	\item the population is split into anti-conformists, or \emph{hipsters}, taking their decisions in opposition to the majority, and \emph{mainstreams} that would rather follow the majority
	\item we explicitly take into account the time needed by each individual to detect and react to changes of states. 
\end{itemize}
This hindrance in the communication and processing of information is generally a realistic feature of social, biological or physical systems. Indeed, a change in the configuration of a physical particule, the firing of a neuron or the issuing of a new fashionable shapka is generally not instantaneously perceived by the collective system and it may take some time to integrate this information for future choices, leading to a delay accounting both for the transmission of information, and the integration of information. Moreover, delays may be distributed heterogeneously between various agents: it may depend upon the proximity of the two agents (physical or more abstract), on the type of agent, and on the geometry of interactions (network structure). Despite their prominence, delays are often neglected in a first approximation in physics or social systems. However, in control theory or in computational neuroscience for instance, delays are known to shape the collective dynamics~\cite{roxin-brunel-etal:05,brunel-hakim:99}. We shall thus investigate in this paper the role of delays in simple statistical physics models. 

Another important aspect we shall consider in the present framework is the fact that interactions between agents are heterogeneous. In other words, specific individuals may have more influence than others, at least to the eyes of some: trend makers, bloggers, editorialists in economical journals, neighbors, friends, to cite a few examples. Here, we consider random interconnections, but with a standard deviation scaling as the inverse of the network size. In that case, there is no impact of the variance of the connectivity between individuals, but the correlation between the delay and the interaction weight play a determinant role. An interesting perspective could be to analyze cases where the standard deviation of the interaction decays much slower, as the inverse of the square root of the network size, in which case transitions will likely arise due to randomness in the connections as observed in neural networks~\cite{sompolinsky1988chaos,hermann2012heterogeneous}.

The toy model we investigate here is thus a simple interacting systems, composed of two caricatural kinds of individuals, the \emph{hipsters}, pure anti-conformist systematically taking their decisions with a tendency to oppose to the majority, and \emph{mainstreams}, that are systematically biased to that follow the majority. We show that the combination of anticonformists and effective delays of interactions consistently induce non-stationary solutions in which no equilibrium is reached in the system and the population keeps oscillating between distinct choices. The paper develops a detailed study of the simplest case whereby individuals make binary choices; we show that when delays in detecting the trend are too large, or when mainstream individuals are the majority, all hipsters align and do the same at the same time. This paper extends our unpublished preprint~\cite{touboul2014hipster} in the direction of asymmetric interactions with rigorous analytics on the bifurcations of the system in the absence of delays. We note that~\cite{touboul2014hipster} has been the basis of subsequent developments in various directions, including investigations of the role of complex networks topologies~\cite{juul2017hipsters}, asymmetric interactions between classes of individuals~\cite{collet2016rhythmic} (that we extend here), or socio-economic applications~\cite{pra2018climb,haidt2017disagreeing,jackson2016market,dairhythmic}. 

The paper is organized as follows. Section~\ref{sec:twotypes} introduces the formalism used in the paper: it provides the definition of the stochastic hipster model and derives the mean-field equations describing the dynamics of the system when the number of interacting individuals diverges. We study the role of delays in this equation in section~\ref{sec:delaytransition}. We identify a delay-induced Hopf bifurcation associated with a sudden synchronization of hipsters when delays are large enough. In section~\ref{sec:asym}, we investigate similar transitions in the absence of delay, induced either by asymmetric interactions or timescales. We discuss those results and perspectives of this study in the conclusion.

\section{The hipster model}\label{sec:twotypes}
We introduce here the basic framework of the study, the network equations, the associated limit equations, and discuss, in a general setting, the stability of the disordered states where hipsters are as distinct as it is possible. 
\subsection{The binary hipster model}
We consider an interacting agents system composed of $n$ individuals whose state $(s_{i})_{i=1\cdots n}$ can take one of two values: $s_{i}\in\{-1,1\}$ for all $i\in \{1,\cdots,n\}$. The state of the network at time $t$ is thus given by a vector $\s(t)=(s_{1}(t),\cdots,s_{n}(t)) \in \{-1,1\}^n$ following a stochastic dynamics, whereby each individual may switch state at any point in time with a rate depending on:
\begin{itemize}
	\item the `trend' they individually perceive at time $t$, denoted $m_{i}(t)$, namely their perception of the imbalance in the states of other agents, and
	\item their conformist (mainstream) or anti-conformist (hipster) nature.
\end{itemize} 
The model simply assumes that agents are heterogeneous. Across the population, the probability of hipster is fixed, and the way one given individual interacts with its environment is idiosyncratic: it depends on the individual in question, although statistically the interaction probability will be considered identical. This heterogeneity leads us to consider a \emph{random environment} for the dynamics: the type of a given individual and the way it interacts with the environment is drawn from a probability distribution, and remains fixed (frozen, or quenched) during the evolution of the network. In addition to this heterogeneity, the system is also \emph{stochastic}: agents take decisions at random times and with a fixed level of randomness. We will consider that each individual switches state at a random time, with an instantaneous rate depending on the trend felt. 

\subsubsection{Random Environment}
Among the population of $n$ individuals, a proportion $q$ is anti-conformists and $p=1-q$ is mainstream. Each individual $i\in \{1,\cdots, n\}$ is thus considered hipster or mainstream independently, with probability $q$ and $p$ respectively. We thus draw, prior to the evolution of the network, a (quenched) sequence of independent identically distributed Bernoulli variables $(\eps_i)_{i=1\cdots n}\in\{-1,1\}^n$ representing the nature of each individual:
\[\eps_{i}=\begin{cases} -1 & \text{(hipster), with probability }\, q \\
1 & \text{(mainstream), otherwise. }\\
\end{cases}\]
Because there are only two possible states, the only element to be defined to complete the description of the Markov model is the rate of change of each individual, which, as indicated above, depends on the trend felt by agent $i$, denoted $m_i(t)$. This trend depends on the previous states of all individuals and the relative weights with which individual $i$ perceives the environment. We consider that individual $i$ assigns a fixed weight $J_{ij}\geq 0$ to the choice of individual $j$: if the style of individual $j$ matters to $i$, $J_{ij}$ will be large, and $J_{ij}=0$ if the style of $j$ has no influence of the future choice of $i$. Moreover, individual perceives and integrates a change in the state of individual $j$ after a delay $\tau_{ij}$. These two elements leads to model the trend seen by individual $i$ at time $t$ as:
\[m_i(t)=\frac 1 n \sum_{j}J_{ij} s_j(t-\tau_{ij})\]
We will make the assumption that for any fixed $i$, the weights and delays pairs $(J_{ij}, \tau_{ij})_{j=1\cdots n}$ are independent random variables with distribution $p_{\eps_i,\eps_j}$ that only depend on the hipster-mainstream nature of $i$ and $j$. This environment is also considered as a quenched disorder: the variables $(J_{ij}, \tau_{ij})_{j=1\cdots n}$ are drawn prior to the stochastic evolution of the network, and are frozen during the evolution in time. An important example that we treat here is when both delays and weights depend on a hidden variable, which is the relative `proximity' of two agents, for instance the distance in space between the two individuals (see section~\ref{sec:binary_ring}). 

Throughout the paper, we shall assume that the delays $\tau_{ij}$ are bounded, and that the connectivity coefficients $J_{ij}$ are regular, having at least a finite second moment. The variables $(\eps_{i})_{i\in\{1\cdots n\}}$ and $(J_{ij},\tau_{ij})_{i,j \in \{1\cdots n\}}$ constitute the random environment of the stochastic evolution. 

\subsubsection{Stochastic Evolution}
\begin{figure}
\begin{center}
\includegraphics[width=.7\textwidth]{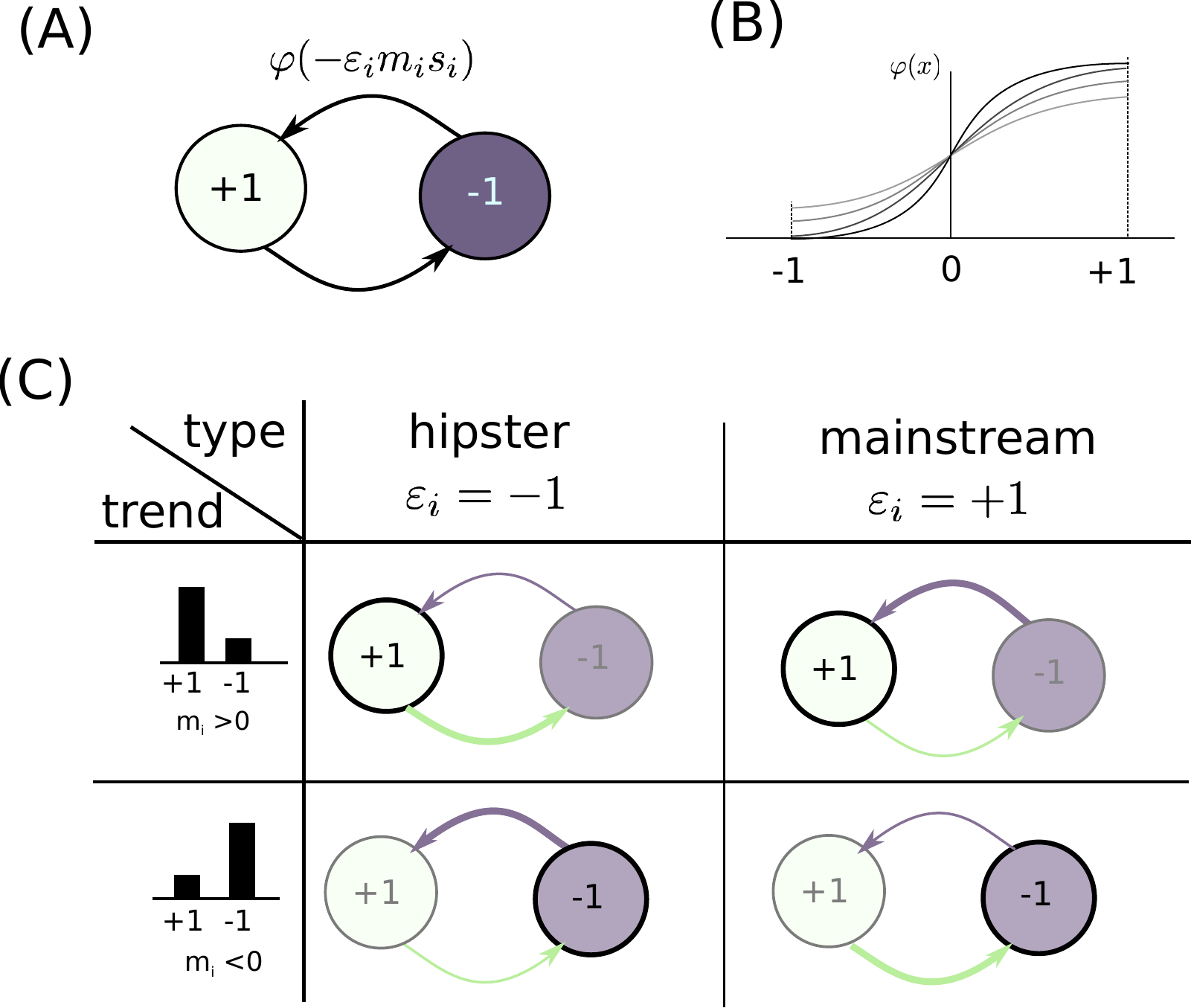}
\end{center}
\caption{The binary hipster model: (A) each individual has a state $s_i=\pm 1$, and switches to the other state depending on the trend felt $m_{i}$ and its hipster nature ($\eps_{i}=-1$) or mainstream ($\eps_{i}=1$) nature. The random transitions occur at a rate $\f(-\eps_{i}s_{i}m_{i})$ where $\f(x)$ is a sigmoidal function (B) depending on a sharpness parameter $\beta$ that account for the level of determinism in the transition: the larger $\beta$, the sharper the transition, and the less probable non-preferred transitions. We depicted distinct possible transition functions with color indicating the value of $\beta$ (darker lines correspond to larger $\beta$). The table in (C) summarizes the 4 possible situations for the two types of individuals and 2 types of majoritary trend.}
\label{fig:binarymodel}
\end{figure}

Once a configuration is fixed, the state of each individual evolves according to a continuous-time Markov jump process. In detail, given the state $\s(t)$ of the network at time $t$, the individual $i$ switches state ($s_i \to -s_i$) with a rate $\f(-\eps_i \,m_i(t)\,s_i)$ where $\f$ is a non-decreasing positive map (a sigmoid function, see Fig.~\ref{fig:binarymodel}). 

In that model, if the state $s_i(t)$ is opposite to the felt trend $m_i(t)$, then the product $m_{i}(t)s_{i}(t)$ is negative, and therefore (see Fig.~\ref{fig:binarymodel}(C)): 
\begin{itemize}
	\item a mainstream ($\eps_i=1$) will feel a strong incentive to switch state (high switching rate), and the more unmitigated the opposite consensus is perceived, the larger $\vert m_{i}(t) \vert$, and thus the larger the switching rate;
	\item hipsters ($\eps_{i}=-1$), on the contrary, will feel compelled to keep their originality, have a low switching rate, lowest when an unmitigated opposite consensus is observed ($\vert m_{i}(t)\vert$ large). 
\end{itemize}	
A variety of choices can be made as of the transition map $\f$. A particularly relevant parameter of the map is its \emph{gain} $\beta=\f'(0)$. {This parameter controls the sensitivity of the individuals to small imbalances around consensus: if $\beta$ is small, a small imbalance around the consensus will lead to a small modification of the switching rate, but for larger gains, a sharper response arises and small imbalances around the consensus state may have strong impact on the switching rate} (see Fig.~\ref{fig:binarymodel}(B)). For this reason, $\beta$ is often considered as a parameter quantifying the level of noise, and is called in thermodynamics the inverse temperature parameter. 

To fix ideas, we shall consider in this manuscript $\f(x)=1+\tanh(\beta x)$. We expect little dependence of the results if choosing another form of sigmoid function with the same gain $\beta$. 

\subsubsection{Comparison to classical models in statistical mechanics}
One of the main {innovations} of the model is the presence of delays, which are rarely considered in canonical statistical mechanics systems. Beyond delays, the fact that agents interact in a totally asymmetric manner makes the system quite distinct from more classical models in the field. In particular, since the impact of the state of individual $i$ on $j$ is not of the same magnitude as the reciprocal action of $j$ on $i$, the dynamics does not derive from a potential and more complex solutions such as periodic or chaotic orbits may arise as in the case of statistical systems in neuroscience. In that sense, our system is comparable to binary neuron models as introduced in early works in the domain~\cite{crisanti-sompolinsky:87b}. Another difference appears in the way we incorporate the mainstream-hipster nature as a characteristic of each individual, which differs from, e.g., the Sherrington-Kirkpatrick spin glass model~\cite{sherrington-kirkpatrick:75} or even canonical randomly connected neural networks~\cite{sompolinsky1988chaos} in which interaction between $i$ and $j$ is assumed to have a random sign. This distinction will actually necessitate a two-dimensional limit that keeps track of the type of individual considered.

\subsection{The large $n$ limit and its transitions}
We now derive the limiting behavior in the large $n$ regime of the switching rates and proportions of individuals of each type in each state. This switching rate is a self-consistent quantity depending on the statistics of the solution that it generates. Fortunately, the limiting system found here is not hard to handle: we will show that the average behavior exactly reduces to a set of deterministic delayed differential equations (DDEs). To study the behaviors, we will thus rely on the well-developed theory of bifurcations of DDEs and uncover transitions related to the delays distribution and other parameters of the system. 

\subsubsection{Thermodynamics limit}
The large $n$ limit of the system can easily be derived using classical physical arguments. In statistical systems, since the impact of one single agent tends to vanish, finite subsets of agents in the large $n$ limit behave independently. Moreover, owing to a law of large numbers, these behaviors also become in the large $n$ limit independent of the environment variables. This \emph{molecular chaos} hypothesis, first formulated by Boltzmann (sto\ss zahlansatz)~\cite{boltzmann} in reference to the hypothesis that the speed of molecules in a gaz prior to collision shall be independent, is termed propagation of chaos property in mathematics~\cite{sznitman1991topics}. Classical theory of jump processes easily generalizes to the present case, and one can follow \emph{mutatis mutandis} the development done in similar systems in the absence of delay~\cite{dai2013curie}. Here, we outline the arguments leading to identify the limit. Under the molecular chaos hypothesis, the interaction term, corresponding to the individually perceived trend:
\[m_i(t)=\frac 1 n \sum_{j}J_{ij} s_j(t-\tau_{ij})=\frac{1}{n} \sum_{j; \eps_{j}=1} J_{ij} s_{j}(t-\tau_{ij}) + \frac 1 n\sum_{j; \eps_{j}=-1} J_{ij} s_{j}(t-\tau_{ij})\]
can thus be considered as the sum of two large-scale sums of independent identically distributed random variables and, owing to a law of large numbers, shall converge towards the sum of the common expectation of each term, namely: 
\begin{equation}\label{eq:m} 
m_{\eps}(t)=\sum_{\eps'=\pm 1} q_{\eps'} \int_{\R^2} j \rho_{\eps'}(t-\tau) dp_{\eps,\eps'}(j,\tau)
\end{equation}
where $\rho_\eps(t):=\E[s_\eps(t)]$ is the averaged value (statistical expectation) of individuals of type $\eps$ at time $t$ and $q_{\pm}$ the proportion of anti-conformists ($q_{-}=q$) and conformists ($q_{+}=1-q$) individuals. 

In that limit, any individual of type $\eps$ will thus switch state according to an inhomogeneous Poisson process with rate $\f(-\eps m_{\eps}(t)s)$, which is indeed a self-consistent equation since the rate depends on $m_{\eps}$, itself depending on the expected value of the solution. Such processes are generally complex; in particular, they do not satisfy the Markov property (even its generalized counterpart taking into account delays) since the jump rate depends on the law of the solution and not on the value of the state only. However, the hipster model enjoys a simple characterization in that limit. Indeed, the thermodynamic limit is univocally described by the two jump rates $m_{\eps}(t)$, that only depend on the knowledge of the average state of individuals in the two populations $\rho_{\eps}$. The Kolmogorov equation associated with this process justifies that the average state in population $\eps$, denoted $\rho_{\eps}$, satisfies the differential equation\footnote{Formally, the Kolmogorov equation is valid only for Markov processes. However, it is not hard to show that the formula extends to the present case, and easy way to derive this equation is to consider the inwards and outwards probability fluxes related to, e.g., state $1$. Denoting $p_{\eps}=\mathbb P[s_{\eps}(t)=1]=\frac 1 2 (\rho_{\eps}+1)$, we obtain:
\[\dot{p}_{\eps}=-p_{\eps} \f(-\eps \rho_{\eps}) + (1-p_{\eps})\f(\eps \rho_{\eps}),\]
which is nothing but equation~\eqref{eq:Kolmogorov}.
}:
\begin{equation}\label{eq:Kolmogorov}
	\dot{\rho}_{\eps}(t)=-2\Big(\rho_\eps(t) + \tanh(-\eps \beta m_{\eps}(t))\Big).
\end{equation}
Equations~\eqref{eq:m} and~\eqref{eq:Kolmogorov} allow a precise analysis of the system and understanding the role of the different parameters. We concentrate on three simple situations in which the role of key parameters are disentangled: (A) a case with constant communication and delay coefficients independent of the individual type, (B) a case where delays and communication coefficients are both dependent on a hidden random parameter, the respective locations of the different individuals, and (C) a model with asymmetric interactions and no delays. Before we proceed, we outline the general methodology used to assess the stability of the disordered state. 

\subsection{Stability of the disordered state}
The above system features, for any parameter set, a completely disordered steady state characterized by an average state for both hipster and mainstream populations equal to $0$, $\rho_{\pm 1}=0$. Indeed, in this situation, the rate of transition of individuals (regardless of their type) from state $1$ to $-1$ is equal to the reciprocal transition rate (rate equal to $\f(0)$), and thus for large $n$, the system remains in a completely disordered at all times. For finite $n$, transient imbalances due to finite-size fluctuations and particularly to the first switches occurring will break the symmetry of the system, may be amplified, in turn leading the system to escape the disordered state and reach other attractors. To study this possibility, we characterize the stability of the disordered solution in the thermodynamic limit. To this end, we consider the linearization of the system about the disordered solution, describing the evolution of a small perturbation $h_{\pm}$ about the disordered state:
\begin{equation}\label{eq:linearized}
\dot{h}_{\eps}(t)=-2\Big(h_\eps(t) -\eps \beta \sum_{\eps'=\pm 1} q_{\eps'} \int_{\R^2} j h_{\eps'}(t-\tau) dp_{\eps,\eps'}(j,\tau)\Big)
\end{equation}
Classically, the spectrum this operator characterizes whether the perturbation $h_{\pm}$ will progressively vanish or amplify: if all eigenvalues of the linearized equation have strictly negative real part, the perturbation vanishes and the disordered state is stable; if at least one eigenvalue has strictly positive real part, the perturbation is initially amplified, and the disordered equilibrium is unstable. Because of the presence of the delay, this equation is infinite-dimensional and the linear operator may feature an infinite number of eigenvalues that may not be well-separated. However, delayed equations have linearized operators in convolutional form~\cite{hale-lunel:93} and it is well-know that the eigenfunctions are exponentials, whose exponents satisfy the so-called \emph{dispersion relationship} that depend on the parameters of the system\footnote{Note that this equation only depends on the fact that the sigmoid allows the disorder state as a solution, and on the gain of the sigmoid at $0$: this observation justifies the statement that the particular form of the sigmoid does not significantly affect the behavior of the system.}. We now derive the stability conditions in various situations.

\section{Delays and synchronization in the hipster model}\label{sec:delaytransition}
In this section, we concentrate on the role of delays in the case of the binary hipster model with symmetric interactions, i.e., where the average impact of the choice of all individuals does not depend on the individual type. We start by studying the simple case of constant delay, before investigating the case of a distributed delay depending on a hidden variable interpreted as the distance between two individuals. 
\subsection{Hipster synchronization in the presence of delays}\label{sec:delays}
Let us start by dealing with situation (A) where $p_{\eps,\eps'}=\delta_{\bar{J},\tau}$. In that homogeneous case, there is no difference in the perception of the trends from hipsters and mainstream viewpoints, i.e. $m_{+1}=m_{-1}$, and these trends are equal to the average state of over the whole population at time $t-\tau$. Denoting:
\[z(t)=q \, \rho_{-1}(t) + (1-q) \,\rho_{+1}(t).\] 
we thus have $m_{\pm 1}(t)=z(t-\tau)$. Using the differential equations~\eqref{eq:Kolmogorov} satisfied by $\rho_{\pm 1}$, we find:
\[\dot{z}(t)=-2 \Big(z(t) + (2q-1)\tanh(\beta \bar{J} z(t-\tau))\Big).\] 
The dynamics of the full $n$-dimensional system in the limit $n\to \infty$ thus reduces to a single delayed-differential equation. The linearized equation around the disordered equilibrium ($z=0$) is given by:
\[\dot{h}(t)=-2 \Big(h(t) + (2q-1)\beta \bar J \; h(t-\tau)\Big).\]
An exponential function $h(t)=e^{\lambda t}$ is an eigenvector of the linearized equation only for $\lambda$ satisfying the \emph{dispersion relationship}:
\begin{equation}\label{eq:dispersion}
\lambda=-2(1+(2q-1) \beta \bar J \; e^{-\lambda \tau}).
\end{equation}
The disordered state is stable if the only complex solutions to the dispersion relationship have a strictly negative real part. This yields specific relationships between parameters that we derive below. 

\paragraph{Stability in the absence of delay}
To understand how delays may modify the dynamics of the system, we first characterize the behavior in the absence of delay (i.e., $\tau=0$). In that case, the dispersion relationship reads:
\[\lambda=-2(1+(2q-1) \beta\bar J),\]
yielding as expected a single eigenvalue (because the system is one-dimensional); the disordered solution is stable if and only if 
\[1+(2q-1)\beta\bar J>0.\] 

A majority of hipsters ($q>1/2$) therefore correspond always to having a stable disordered state. When hipsters form the minority of the population, then it is not hard to show (see e.g.~\cite[Example 3.4.1]{strogatz2018nonlinear}) that the system undergoes a non-degenerate \emph{pitchfork bifurcation} at the point $1+(2q-1)\beta\bar J=0$, a typical exchange of stability bifurcation arising in symmetric systems (here, the equation is equivariant under the reflection symmetry $z\to-z$). At this bifurcation, the disordered state loses stability, and two stable solutions emerge characterizing the emergence of a partial consensus.

Several observations can be made around this result. First, we note that populations in which anticonformists are majority ($q>1/2$) never find consensus, while populations dominated by conformists consensus is found, but only for $\beta \bar J$ large enough, i.e., when randomness in the transition is small enough compared to the impact of the average choices. In detail, consensus are found for $\beta$ larger than a critical value $\beta_c(q)$ that increases with the proportion of hipsters
\[\beta_c(q) =\frac 1 {\bar{J}} \; \frac 1 {1-2q}.\] 
Below this critical noise level, the disordered state $z=0$ loses stability and a state with non-zero trend is found. Heuristically, when there is a majority of anticonformists, these will compensate instantaneously any trend emerging from the mainstream population seeking a consensus, and therefore prevent net trends to establish and maintain. But when there is a majority of mainstreams, a consensus may emerge, but only if the level of randomness in their choices is small enough or if the impact of the global trend ($\bar J$) is large. Hipsters will then consistently oppose to this trend, creating a clear non-trivial hipster trend. The fact that the level of noise at which this equilibrium emerges is lower than that of a pure ferromagnetic spin glass system ($\beta=1$) can be interpreted as the fact that, from a microscopic viewpoint, the systematic frustration and misalignment of hipsters results in an increased effective temperature. 

The the critical transition $q=1/2$ is associated with very complex phenomena. In that regime, in addition to the totally disordered state, a symmetric state stabilizes with hipsters and mainstream individuals locking transiently to distinct states, before switching at seemingly irregular times. To characterize the randomness of switching times, we computed the distribution of duration of time intervals during which one population of individuals (here, hipsters) display a majority of +1 (or -1). The result of long simulations show indeed very irregular distribution of holding times, showing an exponential distribution, implying therefore memoryless switches (parameter $\lambda=0.68$, goodness of fit $0.031$, p-value $<0.001$, test and code from~\cite{clauset2009power}).

\begin{figure}
\includegraphics[width=.8\textwidth]{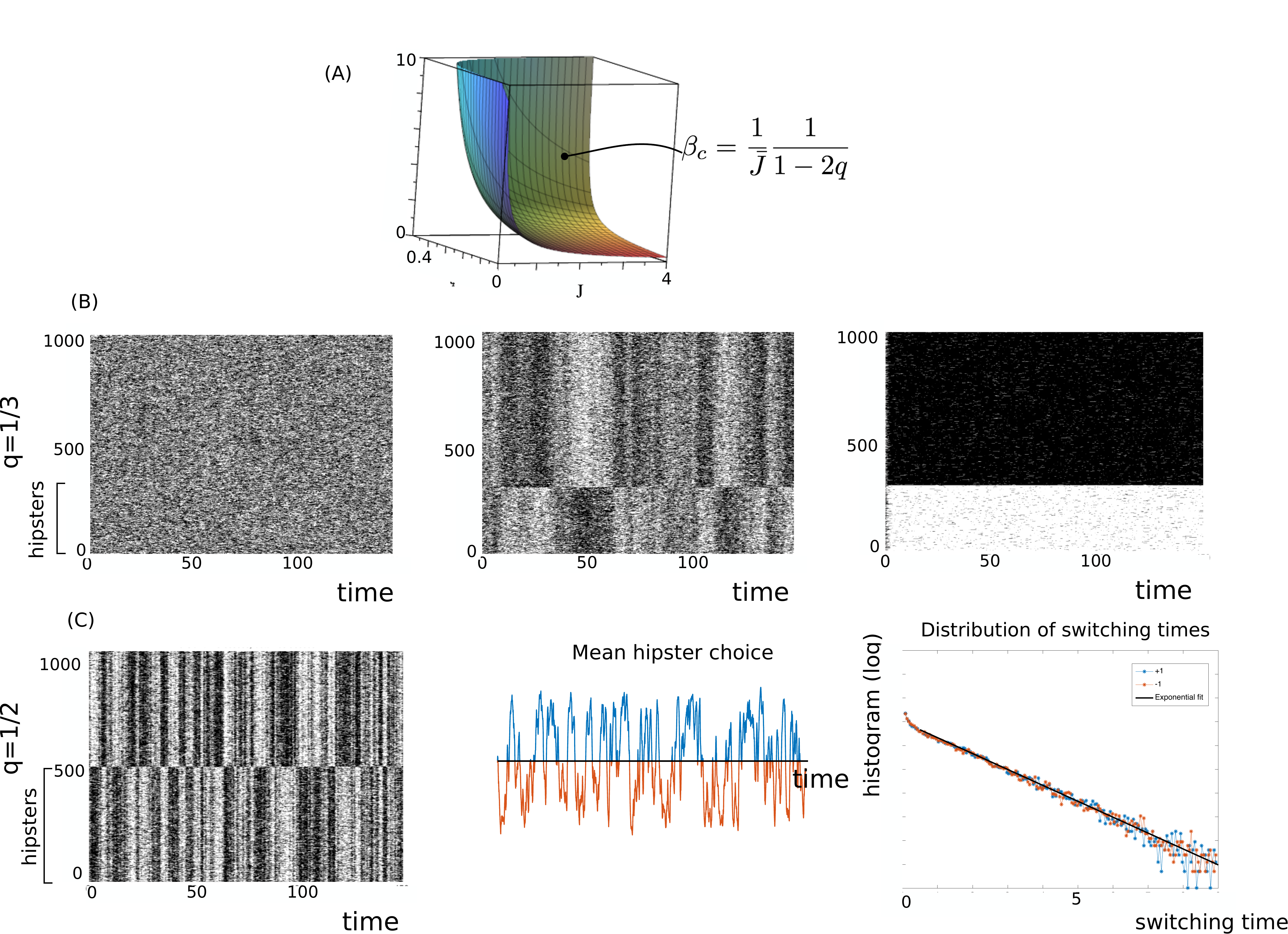}
\caption{Non-delayed system with $\bar J=1$: (A) the critical noise level $\beta_c$ associated with the pitchfork bifurcation of the mean-field system. (B-C) simulations of the stochastic hipster model with $n=1\,000$ agents and no delay. (B): $q=1/3$: subcritical system (left, $\beta=1$) shows a disordered regime, at the critical regime the system switches between distinct disordered states where hipsters and mainstream align transiently (middle, $\beta=\beta_{c}=2.5$), and super-critical system (right, $\beta=5$): the mainstream find a consensus and hipsters oppose to it. (C) $q=1/2$: no bifurcation occurs. At relatively low noise ($\beta=15$), while the average remains close to $0$, the system switches between different states where hipsters and mainstreams are aligned, with switching times exponentially distributed (right).  }
\label{fig:non-delayed}
\end{figure}

\paragraph{Delay-Induced Hopf bifurcation}
 We now study how the presence of the delay affects the stability of the disordered state. We have observed that, in the absence of delay, the disordered state is stable when $\beta<\beta_{c}(q,\bar{J})$. It is easy to show that in this situation, the delay cannot destabilize the disordered state. Indeed, shall there exist $(\beta,\tau)$ for which $0$ is unstable and $\beta<\beta_{c}$, we would then have characteristic roots $\lambda=a+\mathbf{i}b$ with $a>0$ satisfying the dispersion relationship~\eqref{eq:dispersion}. Taking the real part of this relationship, we shall thus have:
\[a=-2 + 2(1-2q)\beta \bar{J} e^{-a\tau}\cos(b\tau)>0,\]
which is impossible since in this situation $\vert 2(2q-1)\beta\bar{J} e^{-a\tau}\cos(b\tau) \vert <2$. 

Therefore, when $q<1/2$, the delay does not induce any disorder. When $q>1/2$, we now show that delays may destabilize the disordered state. To this purpose, we study the characteristic roots solutions of the dispersion relationship~\eqref{eq:dispersion} at the bifurcation point, i.e. when the characteristic roots cross the imaginary axis. We rule out the case where a real eigenvalue crosses the imaginary axis, because such transitions do not depend on the delay and were thus identified in the previous section.  
Pairs of complex conjugated eigenvalues crossing the imaginary axis are given by $\lambda =\pm \mathbf{i}\omega$ where $\omega$ is a solution of equation~\eqref{eq:dispersion}:
\[\mathbf{i}\omega=-2-2(2q-1)\beta\bar{J} e^{-\mathbf{i}\omega\tau}.\]
Denoting $\lambda=(2q-1)\beta\bar{J}>0$, we thus have:
\[2+\mathbf{i}\omega=-2\lambda e^{-\mathbf{i}\omega\tau}.\]
Taking the modulus on both sides and solving for $\omega$ yield $\omega=2\sqrt{\lambda^{2}-1}$, and thus bifurcations only occur when $\lambda>1$.  
Classical angle formulae for complex numbers yields:
\begin{equation}\label{eq:Hopf_Delay}
\tau  = \frac{\pi - \arctan(\sqrt{\lambda^2-1})+2k\pi}{2\sqrt{\lambda^2-1}} \qquad k\in \mathbbm{Z}.
\end{equation}
Therefore, as soon as $\lambda>1$, there exists a maximal value $\tau_{c}(\lambda)$ of the delay for which the disordered solution is stable.

Finding the type of Hopf bifurcation arising in this system characterizes the stability of emerging orbits. The model studied belongs to the general class of equations studied in~\cite{giannakopoulos1999local}. In that paper, using a reduction to normal form of a general nonlinear equation with delay, a simple condition is derived on the nonlinearity that characterizes the type of Hopf bifurcation emerging in these systems. Applying this framework to our case, we define $F(x)=-2(2q-1)\tanh(\beta \bar J x)$, and using the fact that $F''(0)=0$ by symmetry, the result of~\cite{giannakopoulos1999local} characterizes the type of Hopf bifurcation only depending on the sign of $F'''(0)F'(0)$. Here, we have
\[F'''(0)F'(0) = -8 (2q-1)^{2}(\beta\bar J)^{4}<0,\]
implying that the Hopf bifurcation is always supercritical. Therefore, a branch of stable periodic orbits emerges at the instability of the disordered state. 

\begin{figure}
\begin{center}
	\includegraphics[width=.7\textwidth]{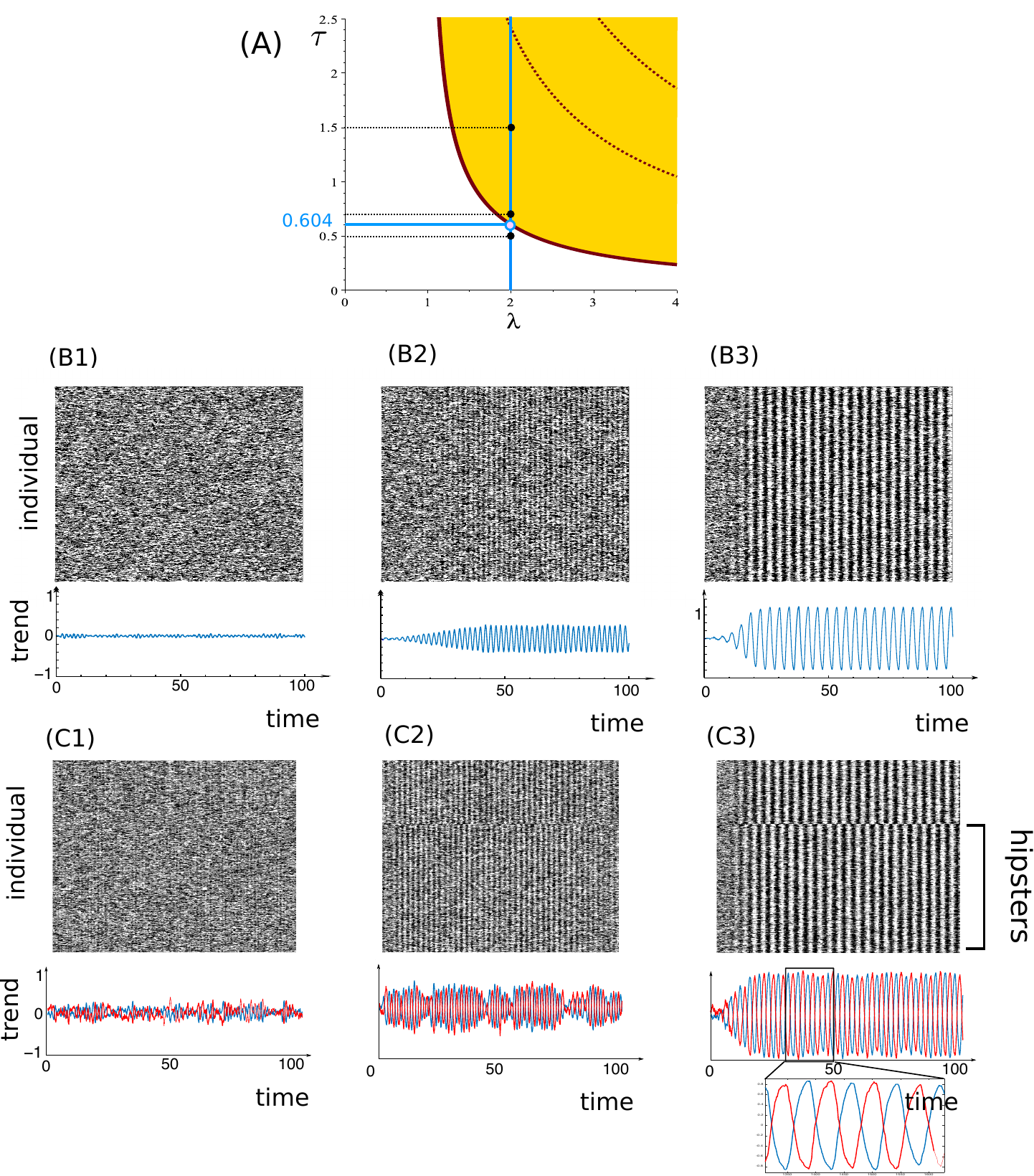}
	\end{center}
	\caption{(A) Delay-induce Hopf bifurcation in the plane $(\lambda,\tau)$.  (B-C) simulations of the discrete system for $n=1\,000$, $\bar{J}=1$, and $\lambda =2$, with distinct parameter combinations: (B) $q=1$ and $\beta=2$, and (C): $q=2/3$ and $\beta=6$. In both cases, the Hopf bifurcation arises at $\tau\approx 0.604$. We depict the behavior of the system for $\tau=0.5$ (B1,C1), $0.7$ (B2,C2) and $1.5$ (B3,C3). Top row: time evolution of all states as a function of time, bottom row: empirical mean for the hipster (blue) or mainstream (red) state.}
	\label{fig:Delay_updown}
\end{figure}

For a delay $\tau$ larger than the smaller positive value of Hopf bifurcations (eq.~\eqref{eq:Hopf_Delay}), a non-trivial periodic solution emerges, oscillating between positive and negative values. The individuals remain synchronized, even if their orientation is not stationary, and switch regularly, in a periodic manner, between positive and negative trends (see Fig.~\ref{fig:Delay_updown})

Heuristically, we can interpret this oscillatory phenomenon as arising from the slowness of the information transmission. Indeed, during the evolution of the network, fluctuations of the trend will tend to be amplified by the delay mechanism. In detail, a random imbalance will be detected after some time and all anticonformist individuals will tend to disalign to this trend, ignoring the fact that an increasing proportion of them do; this will therefore yield the development of a net bias towards the opposite trend, itself later detected and leading to a reciprocal switch. These switches will periodically repeat, yielding synchronized oscillations of the trends. Despite (and actually, in response to) their constant efforts, at all times, anticonformists fail being disaligned with the majority; they actually create the trends they will soon try to escape.

\subsection{Distributed delays and transmission of information}\label{sec:binary_ring}
We consider in this section a more realistic situation where the environment variables $(J_{ij},\tau_{ij})$ depend on a measure of dissimilarity between individual $i$ and $j$. A typical example consists in considering that delays and connections depend on the distance between $i$ and $j$. In that setting, individuals are associated to a hidden position variable $r_i$, which is assumed to take values on a compact set here chosen to be the one-dimensional circle of length $a$, $\S^1_a$. In that setting, individuals communicate after a time proportional to the distance between them, added to a constant delay $\tau_0$ corresponding to the transmission of information $\tau_{ij}=\tau_0+ \vert r_{ij}\vert=:T(r_{ij})$ with $r_{ij}$ is the distance between $i$ and $j$. 

Moreover, the amplitude of the interaction coefficients $J_{ij}$ also depend on $r_{ij}$, modeling the fact that remote individuals have a smaller probability to communicate than nearby individuals. We assume here that the distance affects the probability of a the existence of a link, but not  the amplitude of the connectivity when a link exists, assumed to be equal to a constant value $\bar J>0$. We consider here that an individual communicates with another one located at a distance $r$ with a probability $\psi(r)$ that decays with $r$~\footnote{This is equivalent to an attenuation of the signal with the distance between the two individuals, i.e. to consider communication strength equal to $\bar{J}\psi(r)$.}. 

In that model, the delays are random only due to the randomness of the distance between individuals, and connections are doubly stochastic: they are random variables with parameter depending on the random distance between individuals: $J_{ij} = \bar J \xi_{ij}$ with $\xi_{ij}$ a Bernoulli random variable of parameter $\psi(r_{ij})$. Note that, owing to their common dependence on $r_{ij}$, delays and weights are correlated. Their distribution depends on the distribution of the distances between two points uniformly drawn on the circle, which, owing to the symmetry of the system and periodic boundary conditions of the circle\footnote{Note that if no periodicity is assumed, closed-form formulae can also be obtained. In that case, the distribution has a linearly decaying slope:
$d{\eta}(r)=\left(\frac{2}{a}-\frac{2r}{a^2}\right)\mathbbm{1}_{[0, a]}(r)dr,$ 
and the dispersion relationship can still be obtained explicitly, and the same developments as below are possible, replacing the current expression of $Z(\Omega,\Gamma)$ below by:
 \[Z(\Omega,\Gamma)=- \frac{2(2q-1)\beta \bar{J}}{\Gamma+\mathbf{i}\Omega}\left(1-\frac{1}{\Gamma+\mathbf{i}\Omega} + \frac{e^{-(\Gamma+\mathbf{i}\Omega)}}{\Gamma+\mathbf{i}\Omega}\right)\] 
(see e.g.~\cite{touboul2012limits,quininao2015limits}).}, is the uniform distribution on the interval $[0,a/2]$, and we denote $\eta$ the associated density. 

Using the rule of total probability, we thus obtain that the density of the pair $(J_{ij},\tau_{ij})$ is given by 
\[d p(J=j,\tau)=\begin{cases}
\int_{\S^a} \psi(r) \delta_{\{\tau=T(r)\}} \eta(r) dr & j=\bar{J}\\
\int_{\S^a} (1-\psi(r)) \delta_{\{\tau=T(r)\}} \eta(r) dr & j=0.\\
\end{cases}
\]

Assuming that both mainstream and hipsters have the same probability distributions of delays and coupling coefficients (i.e., $dp_{1,-1}=dp_{-1,1}=dp$), we have $m_{+1}(t)=m_{-1}(t)$, and similarly to the previous case, we can reduce the system to a one-dimensional equation on $z(t)=q \rho_{-1}(t)+(1-q)\rho_{+1}(t)$:
\begin{align*}
\dot{z}(t) &=-2\left(z(t)+(2q-1)\int \tanh(\beta j z(t-s))\;dp(j,s)\right)\\
&=-2 \left( z(t)+(2q-1)\int_{0}^{a/2} \tanh\Big(\beta \bar J z\big(t-T(r) \big) \Big)\psi(r)\;\eta(r) dr\right)\\
\end{align*}

\begin{figure}[h]
	\centering
		\includegraphics[width=.75\textwidth]{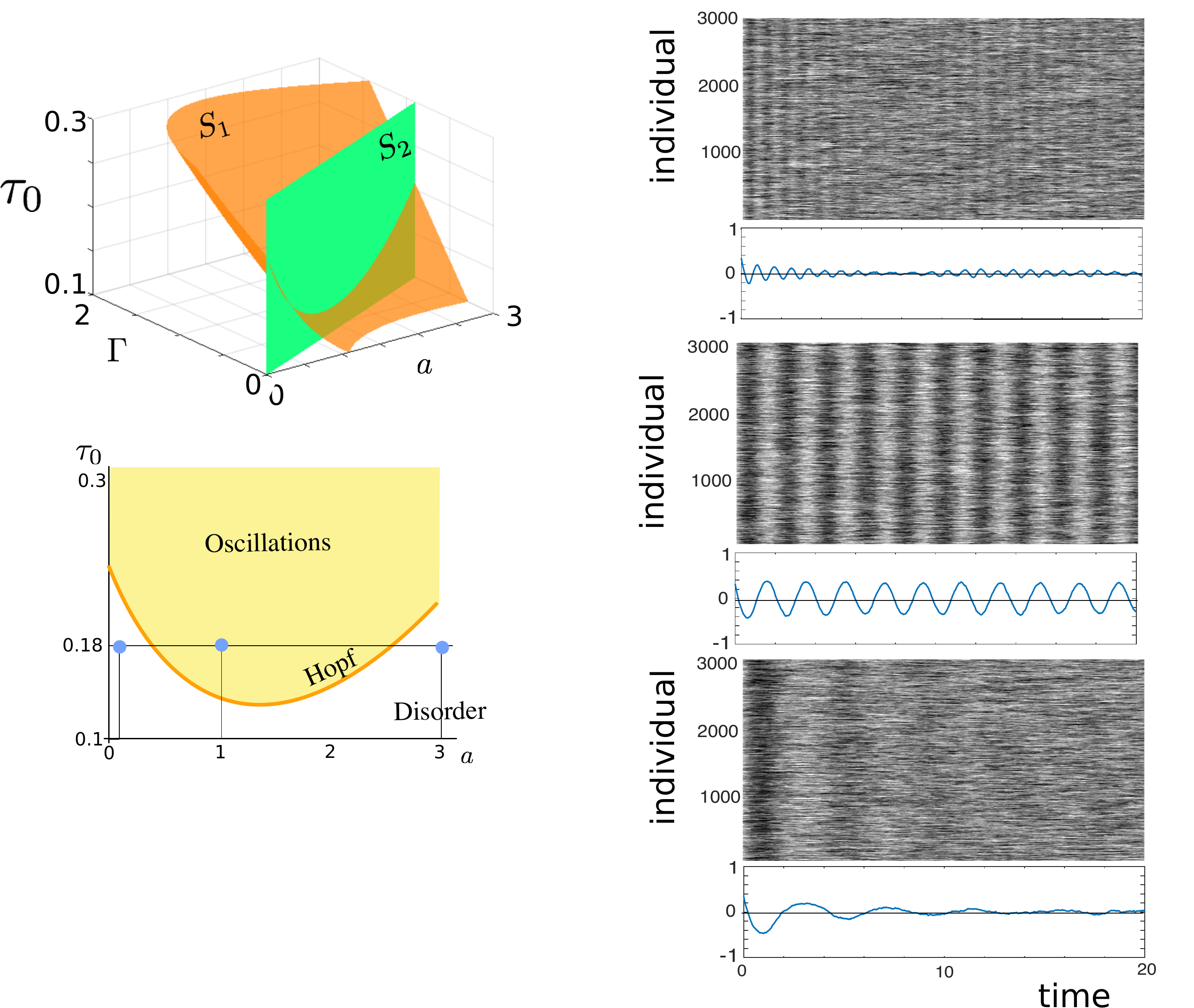}
	\caption{Space-dependent delays and connectivity in a hipster-only situation (see Supplementary Fig.~\ref{fig:Space_q} for the dependence in $q$).: bifurcations as a function of the length $a$ of the interval on which hipsters communicate. Parameters $\beta=4$, $\gamma=0.3$. Left: Computation of the Hopf bifurcation line as a function of $\tau_{0}$ and $a$  (orange line, bottom left) as the intersection of the two surfaces $S_{1}$ (orange) and $S_{2}$ (green surface). For $\tau_0=0.18$, the system shows no synchronization for small intervals (top right, $a=0.1$), synchronization for intermediate intervals (middle right, $a=1$), and no synchronization for large intervals (bottom right, $a=3$), as visible in the simulation of the Markov chain with $N=3\,000$ individuals, together with the computed trend below.}
	\label{fig:Space}
\end{figure}

For $\psi(r)=e^{-\gamma r}$, we can easily compute the linearized operator and derive the equation of the Hopf bifurcation curve in the space of delays and size $a$. Indeed, in that case, the eigenvalues $\xi$ of the linearized operator are solutions of the dispersion relationship:
\[\xi=-2\left (1+(2q-1)\beta \bar J \int_0^{a/2} e^{-(\gamma+\xi)r-\xi\tau_{0}} \,\frac{2\,dr}{a}\right)\] 
which can be integrated explicitly. Hopf bifurcations arise only if one can find parameters of the model, and a positive quantity $\omega>0$, satisfying the relationship:
\begin{equation}\label{eq:OmegaSpace}
	\mathbf{i}\omega = -2 - \frac{4(2q-1)\beta \bar{J}}{a(\gamma+\mathbf{i}\omega)}\left(1-e^{-\frac{a(\gamma+\mathbf{i}\omega)}{2}}\right)e^{-\mathbf{i}\omega\tau_0}.
\end{equation}
This equation cannot be solved in closed form as in the previous case, but however it is easy to express the locus of the Hopf bifurcation in the parameters space $(a,\tau_0)$ as a parametric curve, and therefore access with arbitrary precision the parameters associated with Hopf bifurcations (see Fig.~\ref{fig:Space}). 

Indeed, solving this equation in $(a,\tau_0)$ yields:
 \[\begin{cases}
 	a_{c}(\Omega,\Gamma) &= \Omega \left(\vert Z(\Omega,\Gamma)\vert^2-4\right)^{-1/2}\\
 	\tau_c (\Omega,\Gamma) &= \left(Arg(Z(\Omega,\Gamma)) - Arg(\mathbf{i}\frac{\Omega}{a}+2) + 2k\pi\right)\frac a {\Omega}\\
	\Gamma&=\gamma \,a
 \end{cases}\]
 with $\Omega=a\omega$ and $\Gamma=a\gamma$, and $Z(\Omega,\Gamma)$ the second term of the righthand side of~\eqref{eq:OmegaSpace}:
 \[Z(\Omega,\Gamma)=- \frac{4(2q-1)\beta \bar{J}}{a(\gamma+\mathbf{i}\omega)}\left(1-e^{-\frac{a(\gamma+\mathbf{i}\omega)}{2}}\right).\] 
 While the rescaling of $\omega$ into $\Omega$ is not an issue (as $\omega$ is not a parameter of the equation), the rescaling of $\gamma$ by $a$ is an issue, as we are looking for parameter sets associated with bifurcations of the system. This is why we have added to the system the trivial equation $\Gamma=\gamma\,a$. The set of three equations defines the intersections of two surfaces parameterized by $(\Omega,\Gamma)$ in the three-dimensional space $(a,\gamma, \tau)$ that defines the locus of Hopf bifurcations. Considering $\gamma$ fixed, we thus find the Hopf bifurcation curve as the intersection of two surfaces, $S_{1}$ defined as a parametric surface, and $S_{2}$ a standard surface, in the plane $(a,\Gamma,\tau)$:
 \[S_{1}:\begin{cases} \R\times \R_{+}\mapsto \R^{3}\\
 (\Omega,\Gamma)\mapsto (a_{c}(\Omega,\Gamma),\Gamma,\tau_{c}(\Omega,\Gamma))\end{cases} \qquad S_{2}:\begin{cases} \R_{+}\times \R_{+}\mapsto \R^{3}\\
 (a,\tau)\mapsto (a,\gamma a,\tau)\end{cases}.\]
 The thus obtained surfaces and Hopf bifurcation curve in the plane $(a,\tau_{0})$ are depicted in Fig.~\ref{fig:Space} for a system composed only of hipsters, and in Supplementary Fig.~\ref{fig:Space_q} for various values of $q$. 

Interestingly, this curve displays a non-monotonic shape. It shows that there is an optimal spatial extension of the hipster population most favorable for synchronization: populations spreading on too small or too large intervals will not synchronize, and there exists a specific length interval in which hipsters synchronize. This effect is actually the result of two competing mechanisms: increasing the size of the interval makes the average delay increase (as $a/2$), but the variance of the delays increases as well, which reduces the coherence of the signal received, and may make synchronization harder. 

\section{Asymmetric interactions in the binary hipster model}\label{sec:asym}
In the previous section, we have assumed that the distribution of delays and connectivity coefficients did not depend on the types of individuals. Here, we explore the role of an asymmetry in the connectivity levels between populations. A similar model was discussed in Collet and collaborators~\cite{collet2016rhythmic}, and we come back to this analysis and extend it with an extensive codimension-two bifurcation diagram of the system allowing to identify in detail the regions of the parameters associated with synchrony, and with an interpretation of these results in terms of the interaction coefficients. This leads us in particular to identify a surprising, and somewhat paradoxical new transition leading to synchrony when the randomness of choices exceeds a given threshold. 

\subsection{Theoretical analysis in the absence of delay}\label{sec:theoreticalAsym}
We now assume that the delay of communication between any pair of individuals $i,j$ is equal to $0$, and that the connectivity coefficients $J_{ij}$ are centered at a value that depend on the populations of the two individuals, denoted with a slight abuse of notation $J_{\eps_{i},\eps_{j}}$ . For simplicity, we relabel by $2$ the connectivity indices associated with the hipster population ($\eps_{i}=-1$). The novelty of this model compared to the previous one is that individuals weight differently the states of other individuals depending on the populations they belong to: $J_{11}$ (resp., $J_{21}$) quantifies the importance of the mainstream trend perceived by mainstreams (resp., hipsters) and $J_{12}$ (resp., $J_{22}$) the importance of hipsters choices on the mainstream (resp., hipsters) decisions.  

Contrasting with the symmetric case, the asymmetric interaction case does not reduce to a single equation. Indeed, the Kolmogorov equations associated with that system in the large $n$ limit~\eqref{eq:Kolmogorov} now read:
\begin{equation}\label{eq:Asym}
\begin{cases}
\dot{\rho}_{+1}&=-2 \left[\rho_{+1}+\tanh\left(-\beta \left(J_{11} (1-q) \rho_{+1}+J_{12} q \rho_{-1}\right)\right)\right]\\
\dot{\rho}_{-1}&=-2 \left[\rho_{-1}+\tanh\left(\beta \left(J_{21} (1-q) \rho_{+1}+J_{22} q \rho_{-1}\right)\right)\right]
\end{cases}
\end{equation}
and the Jacobian matrix at the disordered state $\rho_{\pm 1}=0$ is given by:
\begin{equation}\label{JacobianAsym}
A=\left(
\begin{array}{ll}
 -2+2\beta J_{11} (1-q) & 2\beta J_{12} q\\
  -2\beta J_{21} (1-q) &-2-2 \beta J_{22} q
\end{array}
\right).
\end{equation}
Instabilities around the disordered state arise when at least one eigenvalue of the Jacobian matrix crosses the imaginary axis, which can be identified through changes the signs of the determinant and trace of the Jacobian matrix:
\[\begin{cases}
Det:= 4[(1+\beta J_{22}q)(1-\beta J_{11} (1-q)) + \beta^{2} J_{12}J_{21} q (1-q)] \\
Tr:=2[-2+\beta (J_{11}(1-q)-J_{22}q)],
\end{cases}\]
which we rewrite as:
\begin{equation}\label{eq:DetTrace}
\begin{cases}
Det:= 4[-\beta^{2} D -\beta T+1] \\
Tr:=2[-2+\beta T],
\end{cases}
\end{equation}
with 
\[\begin{cases}
T=(J_{11}(1-q) -q J_{22})= (J_{11}- q(J_{11}+J_{22}))\\
D=q(1-q) Det[\mathbf{J}]
\end{cases}\]
where $\mathbf{J}$ denotes the matrix with element $J_{ij}$. The short-hand notation $g_{i1}=(1-q) J_{i1}$ and $g_{i2}=q J_{i2}$ will be use to obtain compact expressions. In these notations, $D=g_{11}g_{22}-g_{21}g_{12}$ and $T=g_{11}-g_{22}$. 

Before we proceed, let us emphasize again the symmetry of the system with respect to the central reflection $(\rho_{1},\rho_{-1})\to(-\rho_{1},-\rho_{-1})$): mainstream and hipsters do not show any bias towards $+1$ or $-1$ states. The bifurcations of the system at the disordered equilibrium will thus reflect that symmetry. 

By the Routh-Hurwirtz criterion~\cite{strogatz2018nonlinear}, we know that the disordered state is stable if $Det>0$ and $Tr<0$, and transitions occurring when one of the conditions is no more true are associated with bifurcations: when $Det=0$ and $Tr<0$, 
a pitchfork bifurcation likely arises, associated with a change of stability of the fixed point, and mathematically this corresponds to the case where one real eigenvalue of the Jacobian matrix crosses the imaginary axis. Reduction to normal form allows assessing the stability on the equilibria emerging from that bifurcation~\cite{kuznetsov2013elements,guckenheimer-holmes:83}. When $Det=0$ and $Tr>0$, a pitchfork bifurcation also arises, yet this bifurcation is not associated with a change of stability of the disordered state (one of the eigenvalues of the Jacobian matrix remains strictly positive across the bifurcation), and the equilibria emerging from that bifurcation are also unstable. Hopf bifurcations arise when $D>0$ and $Tr=0$, in which case two complex conjugate eigenvalues cross the imaginary axis, and a periodic orbit arises, whose location and stability again depend on nonlinear terms.

\begin{proposition}[Pitchfork bifurcations]\label{prop:pitch}
The asymmetric hipster model features:
\begin{itemize}
\item A single pitchfork bifurcation if $D> 0$ at $\beta_{p2}=\frac 1 2 \left(-\frac{T}{D} +\sqrt{\frac{T^{2}}{D^{2}}+\frac 4 D}\right)$;
\item A single pitchfork bifurcation for $D=0$ at $\beta_{c}=1/T$;
\item Two pitchfork bifurcations if $D< 0$ and $T>2\sqrt{-D}$, for $\beta_{p1/p2}=\frac 1 2 \left(-\frac{T}{D} \pm \sqrt{\frac{T^{2}}{D^{2}}+\frac 4 D}\right)$. 
\end{itemize}
In the first two cases, the disordered state is stable if and only if $\beta<\beta_{p2}$ or $\beta<\beta_{c}$ respectively.
In the latter case, the disordered state is stable if and only if $\beta<\beta_{p1}$. 

In the degenerate case $D=0$, the pitchfork bifurcation is supercritical. In the case $D\neq 0$, the type of the pitchfork bifurcation is determined by the sign of $\gamma$:
\[\gamma(\beta)=(-g_{12}g_{21}\beta^{2}+(g_{11}\beta-1)^{2})^{-1}\Big(g_{12}^{3}g_{21}\beta+ (\beta D-g_{22})^{3}(g_{11}\beta-1)\Big)\]
evaluated at $\beta=\beta_{p1}$ or $\beta_{p2}$.
\begin{itemize}
\item For $\gamma>0$, the pitchfork is non-degenerate and subcritical;
\item for $\gamma<0$, the pitchfork is non-degenerate and supercritical.
\end{itemize}
\end{proposition}

\begin{proof}
We break the proof into two parts: A linear analysis where we identify of possible bifurcation points and the stability of the disordered state, and a nonlinear analysis at the bifurcation points consisting of a reduction to normal form to characterize the bifurcation. 

\noindent{\bf Step 1: Linear Analysis. }\\
A real eigenvalue crosses the imaginary axis when the determinant of the Jacobian matrix vanishes, which occurs when the parameters satisfy the relationship:
\[1-\beta T -\beta^{2}D=0\]
Given an interaction matrix $\mathbf{J}$, it is thus easy to find complex values of $\beta$ solutions to this quadratic equation. The only relevant solutions associated with bifurcations are those strictly positive. We distinguish two cases:

\begin{itemize}
\item If the interaction matrix $\mathbf{J}$ is degenerate (i.e., $D=0$, as in the case of identical interaction coefficients), the determinant is linear, and the Jacobian vanishes at $\beta_{c}=1/T$, strictly positive if and only if $T>0$, which can be expressed as a constraint on the proportion of hipsters:
\[q<\frac{J_{11}}{J_{11}+J_{22}}.\] 
At this point, the trace is equal to $-2$, and therefore, for $\beta<\beta_{c}$, both eigenvalues are negative (because the determinant is strictly positive and the trace negative) and thus the disordered state is stable, while for $\beta>\beta_{c}$, the disordered state is unstable (the determinant is negative). 

\item If $\mathbf{J}$ is not degenerate ($D\neq 0$), the determinant of the Jacobian matrix features a single extremum reached at $\beta=-T/2D$, which is a maximum if $D>0$ and a minimum otherwise. Simple analysis based on the value of the determinant at the extremum and the sign of the quadratic coefficient in $D$ allows us to show that:
\begin{itemize}
\item if $D>0$, the system has a unique value of $\beta>0$, given by $\beta_{p2}$, for which the determinant vanishes and changes sign: the determinant is strictly positive if and only if $\beta<\beta_{p2}$, and the disordered state is thus unstable for $\beta>\beta_{p2}$. For $\beta<\beta_{p2}$, the stability of the disordered state depends on the sign of the trace. The trace is an increasing function of $\beta$ vanishing at $\beta_{tr} =2/T$. We have:
\[\frac{\beta_{p2}}{\beta_{tr}}=\frac{T}{2} \cdot \beta_{p2,0} = \frac{T^{2}}{4D}\left(-1+\sqrt{1+\frac{4D}{T^{2}}}\right)\leq 1\]
owing to the properties of the function $x\mapsto x(\sqrt{1+x^{-1}}-1)$. Therefore, $\beta_{p_{2}}<\beta_{tr}$ and the trace is thus negative for any $\beta<\beta_{p2}$.

\item if $D<0$, the determinant vanishes only when $T>2\sqrt{-D}$, condition for the minimum of the determinant to be strictly negative. Therefore, there are exactly two values where the determinant vanishes, $0<\beta_{p1}<-\frac{T}{2D}<\beta_{p2}$, whose expression is classically given as in the statement of the proposition. At these points, the system features pitchfork bifurcations, provided that the non-degeneracy conditions are satisfied (see below). 

The trace of the Jacobian matrix at $\beta_{p2}$ strictly positive, since we have:
\[-2+\beta_{p2} T > -2 +\frac{T^{2}}{2\vert D\vert} > -2 +\frac{4 \vert D \vert}{2\vert D\vert}=0.\]
Since the trace is an increasing function of $\beta$, it is always strictly positive for $\beta>-T/2D$, and thus in particular the disordered state is unstable for $\beta\geq \beta_{p2}$, and no change of stability of the fixed point at this pitchfork bifurcation\footnote{ An alternative method consists in using a a similar argument as the one used for $D>0$ (or for $\beta_{p1}$), analyzing the map $x\mapsto x(1+\sqrt{1-x^{-1}})$ which is always larger than 1.}. 

At $\beta_{p1}$, reasoning as in the case $D>0$, we have
\[\frac{\beta_{p1}}{\beta_{tr}}= \frac{T^{2}}{4\vert D\vert }\left(1-\sqrt{1-\frac{4\vert D\vert}{T^{2}}}\right).\]
The map $x \in [1,\infty) \mapsto x(1-\sqrt{1-x^{-1}})$ is a decreasing map upperbounded by $1$, implying that $\beta_{p1}<\beta_{tr}$, and therefore the disordered fixed point is stable for $\beta<\beta_{p1}$. For $\beta\in (\beta_{p1},\beta_{p2})$, the determinant is negative, implying that the disordered state is unstable. We thus conclude that the disordered state is stable if and only if $\beta<\beta_{p1}$.
\end{itemize}
\end{itemize}

\medskip
\noindent{\bf Step 2: Nonlinear Analysis and Reduction to Normal Form. }\\
To determine the type of the pitchfork bifurcation, we reduce the system to normal form around the bifurcation points identified and compute the cubic coefficient of the system on the one-dimensional center manifold. We start by treating the non-degenerate cases where $D\neq 0$. In that case, we note that the eigenvectors of the Jacobian matrix at $\beta=\beta_{p1}$ or $\beta_{p2}$ are the following:
\[ 0, \;\; V=\binom{- g_{12}\beta}{g_{11}\beta-1}\qquad;\qquad 2T\beta-4, \;\; \binom{-g_{12}\beta }{g_{22}\beta+1},\]
and those of the transpose of the Jacobian matrix:
\[ 0, \;\; V'=N \binom{g_{21}\beta}{g_{11}\beta-1}\qquad;\qquad 2T\beta-4, \;\; \binom{g_{21}\beta }{g_{22}\beta+1},\]
with $N$ the normalizing constant ensuring $\langle V,V'\rangle=1$:
\[N= (-g_{12}g_{21}\beta^{2}+(g_{11}\beta-1)^{2})^{-1},\] 
whose sign is a priori undetermined. 
Following~\cite[Chap. 5.4.2]{kuznetsov2013elements}, the vectors $V$ and $V'$ allow directly accessing the normal form of the system at this point. In the vicinity of the disordered state, the system can be expanded as:
\[\dot x = A x + C(x,x,x) + o(\vert x\vert^{3}),\] 
where $C$ is a trilinear function. Note in that expansion the absence of quadratic terms, related to the symmetry of the system. The restriction of the system to the center manifold, noting the coordinate on the one dimensional manifold $u$, takes the form:
\[\dot{u}=\langle V', C(V,V,V)\rangle u^{3}+o(\vert u \vert^{3})\]
and thus the type of the pitchfork bifurcation is determined by the sign of $\langle V', C(V,V,V)\rangle$. 
Using the Taylor expansion of $\tanh$ at $0$ yields:
\[C(x,x,x)=-\frac{\beta^{3}}{3}\binom{(g_{11} x_{1} + g_{12}x_{2})^{3}}{-(g_{21} x_{1} + g_{22}x_{2})^{3}}\]
so that:
\[C(V,V,V)=\frac{\beta^{3}}{3}\binom{g_{12}^{3}}{(\beta D-g_{22})^{3}}\]
and the type of the pitchfork is thus determined by:
\[\gamma=N(g_{12}^{3}g_{21}\beta+ (\beta D-g_{22})^{3}(g_{11}\beta-1)).\]

For $D=0$, the eigenvectors of the Jacobian matrix are given by:
\[ 0, \;\; V=\binom{- g_{12}}{g_{22}}\qquad;\qquad -2, \;\; \binom{-g_{12}}{g_{11}},\]
and those of the transpose of the Jacobian matrix:
\[ 0, \;\; V'=N \binom{g_{11}}{g_{12}}\qquad;\qquad -2, \;\; \binom{g_{22}}{g_{12}},\]
with $N=-(g_{12} T)^{-1}$. It is easy to check that: 
\[C(V,V,V)=-\frac{\beta^{3}}{3}\binom{(-g_{11} g_{12} + g_{12}g_{22})^{3}}{(g_{21} g_{12} - g_{22}^{2})^{3}}=\frac{(\beta T)^{3}}{3}\binom{g_{12}^{3}}{g_{22}^{3}},\]
and thus the stability depends on the coefficient:
\[\langle V',C(V,V,V)\rangle = -(g_{12} T)^{-1} (g_{11}g_{12}^{3}+g_{12}g_{22}^{3}),\]
and since $T>0$ for relevant bifurcations points (we are looking for positive $\beta$ and, at this bifurcation, $\beta=1/T$), the bifurcation is always supercritical.
\end{proof}

The pitchfork bifurcations are instabilities giving rise to the emergence of new steady states, and are thus not associated with synchronization. We now consider the emergence of Hopf bifurcations, generally associated with oscillatory behaviors. 

\begin{proposition}[Hopf bifurcations]\label{prop:hopf}
For $D<0$ and $T<2\sqrt{-D}$, the system features a Hopf bifurcation at $\beta_{h} = \frac 2 T$, with first Lyapunov exponent:
\[l_{1}=\frac { -2 D \left( g_{11}\,g_{12}-g_{21}\,g_{22} \right) }{ g_{21}\omega^{2} T^{3}}.\]
The disordered equilibrium is stable for $\beta<\beta_{h}$ and unstable for $\beta>\beta_{h}$. Moreover, 
\begin{itemize}
\item if $g_{11}\,g_{12}<g_{21}\,g_{22}$, the bifurcation is supercritical, associated with attractive cycles 
\item if $g_{11}\,g_{12}>g_{21}\,g_{22}$, the bifurcation is subcritical, associated with repulsive cycles.
\end{itemize}
\end{proposition}

\begin{proof}
Hopf bifurcations correspond to instabilities of the system associated with a pair of complex eigenvalues crossing the imaginary axis. We locate these bifurcations and characterize, using nonlinear methods, the type of Hopf bifurcation occurring. 

\noindent {\bf Linear Stability Analysis.}\\
Hopf bifurcation occur when the trace of the Jacobian matrix vanishes while its determinant is strictly positive. Since the trace is a linear increasing function of $\beta$, for any combination of interaction coefficients the trace vanishes for:
\[\beta_{h}=\frac{2}{T}.\] 
Again, this bifurcation is relevant from the system's viewpoint only when $\beta>0$, i.e. when $T>0$, and we discuss this condition below. We note that the parameter $T$ only depends on the self-interaction coefficients $J_{11}$ and $J_{22}$, and the above relationship thus does not impose any constraint on the cross-population interaction coefficients $J_{12}$ and $J_{21}$. Since the determinant is a strictly increasing function of the product $J_{12}J_{21}$, the system will systematically feature a Hopf bifurcation when the product of cross-population interaction terms is large enough. In detail, at $\beta=\beta_{h}$, the determinant simplifies to:
\[Det= -1-\beta_{h}^{2} D = -1-\frac{4D}{T^{2}},\]
and the positivity condition on the determinant is simply $ T<2\sqrt{-D}$. 

For $\beta>\beta_{h}$, the trace is strictly positive, indicating that the disordered equilibrium is unstable for low noise. For $\beta<\beta_{h}$, the trace is negative, and the determinant does not vanish (since the only zeros of the determinant were found for $T>2\sqrt{-D}$ in proposition~\ref{prop:pitch}), so the disordered equilibrium is stable for $\beta<\beta_{h}$. 

\noindent {\bf Nonlinear Analysis and Reduction to Normal Form.}\\
Let us now reduce the system to normal form to assess possible degeneracies of this bifurcation, and the stability of the cycles associated. We proceed classically by considering the Taylor expansion of the system near the disordered equilibrium (see e.g.~\cite[Chap.3]{kuznetsov2013elements}) writing this expansion in the axis given by the eigenvectors or the Jacobian matrix. Up to third order, the system~\eqref{eq:Asym} is given by:
\[\dot{\zeta} = A\zeta + \frac{1}{6} C(\zeta,\zeta,\zeta) + o(\Vert \zeta\Vert^{3})\]
with $\zeta=\binom{x}{y}$, $A$ is the Jacobian matrix given by~\eqref{JacobianAsym} and $C$ is the tri-linear function:
\[C(\zeta_{1},\zeta_{2},\zeta_{3})=  4\beta^{3}\left[ \begin {array}{l} - \Big(g_{11}^{3} x_1 x_2 x_3+ g_{12} g_{11}^{2} \left( x_1 x_2 y_3 + x_1 x_3 y_2 + x_2 x_3 y_1 \right) \\
\qquad \qquad +g_{12}^{2}g_{11}
 \left( x_1y_2y_3+x_2 y_1 y_3+ x_3 y_1 y_2 \right) +g_{12}^{3} y_1y_2y_3\Big) \\
g_{21}^{3} x_1 x_2 x_3 + g_{22} g_{21}^{2} \, \left( x_1 x_2 y_3+x_1 x_3 y_2+x_2 x_3 y_1 \right) \\
\qquad \qquad + g_{22}^{2} g_{21} \left( x_1 y_2 y_3 + x_2 y_1 y_3 + x_3 y_1 y_2 \right) + g_{22}^{3} y_1 y_2 y_3\end {array} \right] 
\]
with $\zeta_{i}=\binom{x_{i}}{y_{i}}$ for $i=1,2,3$. The first Lyapunov coefficient associated with the normal form is given by:
\[l_{1}=\frac{1}{2\omega^{2}} \Re(\omega \langle p, C(q,q,q^{*})\rangle)\]
with $\omega=\sqrt{-1-\frac{4D}{T^{2}}}$ the imaginary part of the eigenvalue at the Hopf bifurcation, and $(p,q)$ the left and right eigenvectors of the Jacobian matrix associated respectively with $-\mathbf{i}\omega$ and $\mathbf{i}\omega$ with $\langle p,q\rangle=1$ and $\langle \cdot, \cdot\rangle$ the complex scalar product. 

It is easy to check that these eigenvectors are given by:
\[q=\binom{1}{\frac{-2g_{21}}{g_{11}+g_{22}+\mathbf{i} \omega T}}\qquad \text{and}\qquad p=\frac{1}{2\omega T}\binom{-2\mathbf{i} g_{21}}{\omega T-\mathbf{i} (g_{11}+g_{22}) }\]
where we took into account the normalization condition $\langle p, q\rangle =1$. The expression of the first Lyapunov exponent dramatically simplifies and we obtain:
\[l_{1} = -2\,{\frac { \left( g_{11}\,g_{12}-g_{21}\,g_{22} \right) 
 \left( g_{11}\,g_{22}-g_{12}\,g_{21} \right) }{ \left( {
\it g11}-g_{22} \right) g_{21}\, \left( {g_{11}}^{2}+2\,{\it 
g11}\,g_{22}-4\,g_{12}\,g_{21}+{g_{22}}^{2} \right) }} = 
\frac { -2 D \left( g_{11}\,g_{12}-g_{21}\,g_{22} \right) }{ g_{21}\omega^{2} T^{3}}.
\]
Since $D<0$ and $T>0$, the first Lyapunov exponent has the sign of $g_{11}g_{12}-g_{21}g_{22}$. 
\end{proof}

The above explicit analysis identified the points where the system features generic bifurcations, and also point towards the existence of codimension-two bifurcations. We summarize in the following proposition those bifurcations and, for some of those, derive the associated normal form. 
\begin{proposition}[Codimension-two bifurcations]\label{pro:codimensiontwo}
The above derived bifurcations meet or degenerate at, at least, three codimension-two bifurcations:
\begin{enumerate}
\item A degenerate pitchfork bifurcation for $D>0$ on the one hand, and $D<0$, $T>2\sqrt{-D}$ on the other hand, and $\gamma=0$. This bifurcation has the normal form:
\[\dot{u}=\lambda_{1} u + \gamma u^{3}+ \eps u^{5}\]
when $\gamma_{2}\neq 0$ and $\eps=sign(\gamma_{2})$, for:
\[\gamma_{2}=\frac{2\beta^{5}}{15} (-g_{12}g_{21}\beta^{2}+(g_{11}\beta-1)^{2})^{-1} (\beta g_{21}g_{12}^{5} + (g_{11}\beta-1)(\beta D-g_{22})^{5}).\]
\item A Bautin bifurcation for $D<0$, $T \in (0,2\sqrt{-D})$, at $\beta_{h}=\frac 2 T$ and $g_{11}\,g_{12}=g_{21}\,g_{22}$, which is non-degenerate when
\[2g_{12} \vert g_{11}+g_{12}q_{2}\vert^{4} - \vert g_{21} + g_{22}q_{2}\vert^{4}\neq 0\]
where $q_{2}=\frac{-2g_{21}}{g_{11}+g_{22}+\mathbf{i} \omega T}$ and $\omega$ the imaginary part of the eigenvalue at the bifurcation. 
\item A double-zero singularity corresponding to a degenerate codimension-two point at $D< 0$ and $T=2\sqrt{-D}$, at $\beta=\beta_{h}(=\beta_{p1}=\beta_{2})$, where a five-branch pitchfork and a Hopf bifurcation simultaneously arise. 
\end{enumerate}
\end{proposition}
Note that the above conditions are not necessarily exclusive, and higher codimension bifurcation points may exist for specific combinations of parameters, that are not studied here. Evidences of the presence such higher-condimension bifurcations is found in Fig.~\ref{fig:TheoreticalCurves}.

The double-zero singularity (third bullet point in Proposition~\ref{pro:codimensiontwo}) is of Takens-Bogdanov type, but appears nonclassical, and its detailed unfolding is not in the scope of the paper. We however undertake in the proof the reduction to normal form, and indicate the possible sources of degeneracy, and provide a numerical investigation of the unfolding of the bifurcation in Figure~\ref{fig:codim2}, where, in addition to the bifurcation curves unfolding from symmetric Takens-Bogdanov bifurcations (see~\cite[Chap. 7.3, p 371--376]{guckenheimer-holmes:83}), a saddle-node bifurcation arises, associated with non-disordered equilibria. 

\begin{proof}

(1). Propositions~\ref{prop:pitch} identified in the case $D<0$, $T>2\sqrt{-D}$ pitchfork bifurcations whose cubic term on the center manifold is proportional to $\gamma$. The value of $\gamma$ may vanish and change sign, at which point the equation of the system at the pitchfork point on the center manifold is no more cubic. To assess the stability of the disordered state at this point, one needs to extend the analysis to derive the fifth-order term in the expansion of the system on the center manifold, and confirm that this coefficient is non-zero. The methodology developed for Proposition~\ref{prop:pitch} extends to the computation of the fifth order coefficient of the vector field along the center manifold. The Taylor expansion of the system up to order 5 can be written as:
\[
\dot x = A x + C(x,x,x) + C_{2}(x,x,x,x,x) +o(\vert x\vert^{5}),
\]
where $x=\binom{x_{1}}{x_{2}}$ and $C_{2}$ is the multilinear map associated with the Taylor expansion of the system; we have: 
\begin{equation}\label{eq:C2}
C_{2}(x,x,x,x,x)=\frac{2\beta^{5}}{15}\binom{(g_{11} x_{1} + g_{12}x_{2})^{5}}{-(g_{21} x_{1} + g_{22}x_{2})^{5}}.
\end{equation}
The dynamics on the center manifold takes the form:
\[\dot{u}=\gamma u^{3}+ \langle V', D(V)\rangle u^{5}+ o(\vert u \vert^{5})\]
with $V$ and $V'$ the eigenvectors of, respectively, the Jacobian matrix and its transpose, associated with the eigenvalue $0$. Using the expressions derived in Proposition~\ref{prop:pitch}, we thus conclude that the fifth order term coefficient, $\gamma_{2}=\langle V', D(V)\rangle$, which is given by:
\[\gamma_{2}=N\frac{2\beta^{5}}{15} (\beta g_{21}g_{12}^{5} + (g_{11}\beta-1)(\beta D-g_{22})^{5}),\]
proving the first point of the proposition.

(2). Proposition~\ref{prop:hopf} identified two points where Hopf bifurcations either disappear or change type, and that thus may constitute the locus of higher-order degeneracies: (a) the case $D=0$, and (b) the point where the Lyapunov exponent $l_{1}$ vanishes. We start considering the case (b), generally associated with the presence of Bautin bifurcations. The case (a) is covered by the point (3.) of the proposition discussed below, and we thus focus on case (b). To confirm the presence of a non-degenerate Bautin bifurcation when $l_{1}=0$, we compute the fifth-order term of the vector field on the center manifold, which would essentially follow the same lines as the derivation of the first Lyapunov exponent. Using the symmetry of the flow and thus the fact that the Taylor expansion of the vector field has no even order terms simplifies drastically formula of the second Lyapunov exponent provided in~\cite[(8.23)]{kuznetsov2013elements}, and we obtain:
\[\l_{2}= \frac{1}{\omega} \text{Re}(g_{32}) - \frac{1}{\omega^{2}} \text{Im}(g_{30}g_{12})\]
where $g_{30}=\langle p,C(q,q,q)\rangle$, $g_{1,2}=\langle p,C(q,q^{*},q^{*})\rangle = 0$ and $g_{32}=\langle p, C_{2}(q,q,q,q^{*},q^{*})\rangle$ for $(q,p)$ the eigenvectors of the Jacobian and its transpose derived in proposition~\ref{prop:hopf}, and $C_{2}$ is the multilinear map associated with~\eqref{eq:C2}. Since at the Bautin bifurcation $g_{30}=0$, $l_{2}$ is equal to the first term of the above formula only, and using the explicit expressions of the eigenvectors derived in proposition~\ref{prop:hopf}, we obtain:
\begin{align*}
\l_{2} &= \frac{1}{2\omega^{2}T} \text{Re}\Big(2\mathbf{i}g_{21} \vert g_{11}+g_{12}q_{2}\vert^{4} (g_{11}+g_{12}q_{2}) - (\omega T+\mathbf{i} (g_{11}+g_{22}))\vert g_{21} + g_{22}q_{2}\vert^{4}(g_{21} + g_{22}q_{2})\Big) \\
&= \frac{g_{21}}{2\omega \vert g_{11}+g_{22}+\mathbf{i} \omega T\vert^{2}} \Big(2g_{12} \vert g_{11}+g_{12}q_{2}\vert^{4} ) - \vert g_{21} + g_{22}q_{2}\vert^{4}\Big), 
\end{align*}
proving the second point of the proposition. 

(3). Eventually, we noticed that $D=0$ splits the parameter space into a regime with a single pitchfork bifurcation from the case with a pair of pitchfork bifurcations, and the singular case where $D<0$ and $T=2\sqrt{-D}$, where two boundaries are met: at this point, the two pitchfork bifurcations collide and disappear, and the Hopf bifurcation disappears at reaching this point. For the pitchfork bifurcations, the transition occurring at $D=0$ smoothly connects the cases $D>0$ and $D<0$. Indeed, we notice that the value of the pitchfork bifurcation $\beta_{p2}$ (having an identical expression in the case $D>0$ and $D<0$) converge towards $\beta_{c}=1/T$. Moreover, the constraint $T>2\sqrt{-D}$ for $D<0$ degenerates to a trivial condition $T>0$. The new pitchfork bifurcation point $\beta_{p1}$ diverges as $-T/D$ when $D\to 0$.

At $T=2\sqrt{-D}$ for $D<0$, the two pitchfork bifurcations collide and disappear, and a Hopf bifurcation emerges. This highly degenerate point probably arises in the present system due to its symmetry. We outline the reduction to normal form of the system at this point. Since that point constitutes a double-zero nilpotent singularity, we follow the methodology of~\cite[Chap. 8.4]{kuznetsov2013elements} but apply it at order three instead of order two, since because of the symmetry of the system terms of even order vanish (see also~\cite[Chap. 7.3]{guckenheimer-holmes:83}). Reducing the Jacobian matrix $A$ into its Jordan form, we derive four vectors $q_{0},q_{1},p_{0},p_{1}$ such that:
\[Aq_{0}= A^{T} p_{1}=0, \; A q_{1}=q_{0} \text{ and } A^{T} p_{0}=p_{1} \]
and moreover
\[\langle q_{0},p_{0}\rangle = \langle q_{1},p_{1}\rangle = 1, \qquad \langle q_{0},p_{1}\rangle = \langle q_{1},p_{0}\rangle = 0.\]
After some simple algebra, we find:
\[q_{0}= \binom{1}{-\frac{2g_{21}}{g_{11}+g_{22}}} \qquad q_{1}=\binom{\frac{T}{2(g_{11}+g_{22})}}{0} \qquad p_{0}=\binom{0}{-\frac{2g_{12}}{g_{11}+g_{22}}}, \qquad p_{1}=\binom{\frac{2(g_{11}+g_{22})}{T}}{\frac{4g_{12}}{T}}.\]
We can thus select $(q_{0},q_{1})$ as a basis and rewrite the equation in the new coordinates, and obtain:
\[\binom{\dot{y_{1}}}{\dot{y_{2}}} = \left(\begin{array}{ll} 0 & 1\\0&0\end{array}\right)\binom{y_{1}}{y_{2}} + \binom{\langle p_{0},F(y_{1}q_{0}+y_{2}q_{1})\rangle}{\langle p_{1},F(y_{1}q_{0}+y_{2}q_{1})\rangle}\]
where 
\[F(x,y)=\binom{2\tanh(\beta(g_{11}x+g_{12}y))-2\beta(g_{11}x+g_{12}y)}{-2\tanh(\beta(g_{21}x+g_{22}y))+\beta(g_{21}x+g_{22}y)},\]
yielding the system:
\[
\begin{cases}
\dot y_{1} = y_{2} - \gamma (\alpha_{1} y_{1}+\delta_{1}y_{2})^{3} + R_{1}(y)\\
\dot y_{2} = \frac{2(g_{11}+g_{22})}{T} \left(-\frac{\beta^{3}}{3} (\alpha_{2} y_{1}+\delta_{2}y_{2})^{3} + \gamma (\alpha_{1} y_{1}+\delta_{1}y_{2})^{3}\right)+R_{2}(y)
\end{cases}\]
with $R_{1}$ and $R_{2}$ two terms of order $4$ or higher, $\gamma=4\beta^{3}g_{12}/3(g_{11}+g_{22})$, $\alpha_{1}=g_{21}-2g_{22}g_{21}/(g_{11}+g_{22})$, $\delta=g_{21}T/2(g_{11}+g_{22})$, $\alpha_{1}=2\alpha(g_{11}+g_{22})/T$ and $\delta_{1}=2\delta(g_{11}+g_{22})/T$.

 To put the system in a more standard normal form, we define $z_{2}=y_{2} - \gamma (\alpha_{1} y_{1}+\delta_{1}y_{2})^{3}$. Up to third order in $y$, we thus obtain:
\[
\begin{cases}
\dot y_{1} = z_{2}\\
\dot z_{2} = A y_{1}^{3}+B y_{1}^{2}z_{2}+C y_{1}z_{2}^{2}+D z_{2}^{3}+R_{3}
\end{cases}\]
with 
\[
\begin{cases}
A &=  \frac{2(g_{11}+g_{22})}{T} (-\frac{\beta^{3}}{3}\alpha_{2}^{3}+\gamma\alpha_{1}^{3})\\
B &=  \frac{2(g_{11}+g_{22})}{T} (-\beta^{3} \alpha_{2}^{2}\delta_{2}+2\gamma\alpha_{1}^{2}\delta_{1}+\gamma(\delta_{1}-3)\alpha^{2})\\
C &=  \frac{2(g_{11}+g_{22})}{T} (-\beta^{3}\alpha_{2}\delta_{2}^{2}+\gamma \alpha_{1}\delta_{1}^{2}+2\gamma (\delta_{1}-3)\alpha_{1}\delta_{1})\\
D &=  \frac{2(g_{11}+g_{22})}{T} (-\frac{\beta^{3}}{3}\delta_{2}^{3}+\gamma(\delta_{1}-3)\delta_{1}^{2}),
\end{cases}\]
and $R_3$ is the rest, negligible compared to the cubic terms highlighted. 
Given the complexity of these terms, it is hard to push one step further the reduction to normal form. In general, a change of time of type $dt=(1+\theta_{1}y_{1}z_{2}+\lambda z_{2}^{2})$ allows, using similar procedure as before, canceling the dependence of the second equation with a quadratic and cubic dependence in $z_{2}$, under a few non-degeneracy assumptions on the coefficients. One then obtains the equation:
\[
\begin{cases}
\dot v_{1} = v_{2}\\
\dot v_{2} = A_{0}v_{1}+A_{1} v_{2} + B_{1}v_{1}^{3}+B_{2} v_{1}^{2}v_{2},
\end{cases}\]
which is the normal form of the Bogdanov-Takens bifurcation in symmetric systems (see~\cite{carr}), and the signs of $B_{1}$ and $B_{2}$ determine the unfolding of the bifurcation. Here, assuming non-degeneracy for the above procedure, one derives extremely complex formulae for $B_{1}$ and $B_{2}$ (using e.g., a formal calculation software as Maple), and studying analytically their sign is very intricate. We thus rely on numerical analysis in Figure~\ref{fig:codim2} to study their unfolding. The unfolding shares a number of common properties with the classical cases of Takens-Bogdanov singularities as well described in~\cite[Chap. 7.3, p 371--376]{guckenheimer-holmes:83}, except the presence of an additional saddle-node bifurcation in the unfolding, associated with the presence of a five-branch pitchfork bifurcation. This inconsistency may be either due to a failure of the non-degeneracy conditions associated with the last transformation in time, or to a vanishing of the coefficients. The formal expressions derived did not allow identifying at this stage which situation occurs here. 
\end{proof}

\subsection{Numerical simulations and interpretations of the instabilities}
\begin{figure}[h]
\includegraphics[width=.9\textwidth]{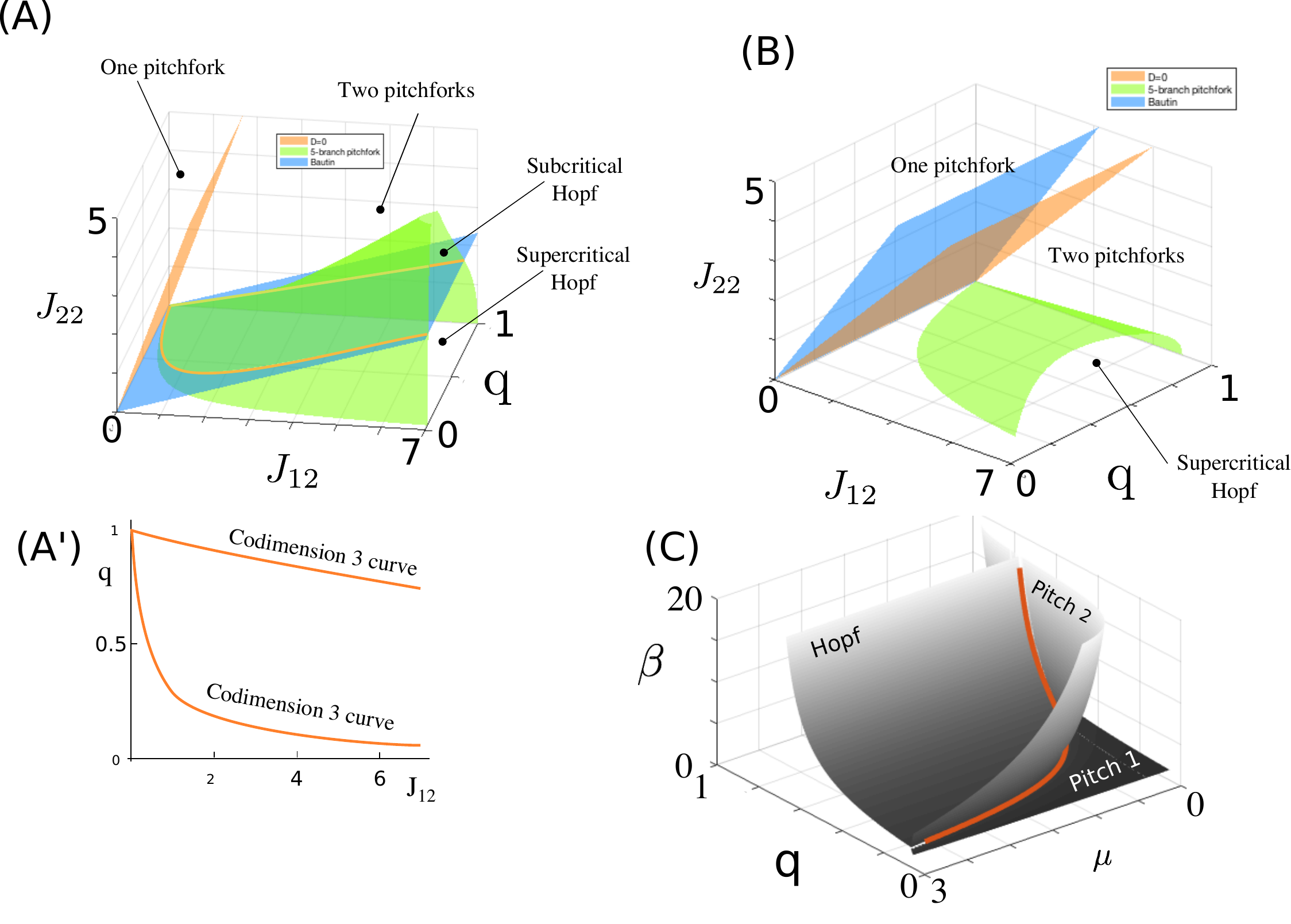}
\caption{Analytical surfaces associated with transitions between the different regimes theoretically derived in section~\ref{sec:theoreticalAsym} (A,B), and bifurcation surfaces (C). (A-B): orange plane (independent of $q$) shows the parameters associated with $D=0$, and corresponds to the switch from regimes associated with one pitchfork bifurcation to regimes associated with two pitchfork bifurcations (see proposition~\ref{prop:pitch}). The 5-branch pitchfork bifurcation at $T=2\sqrt{-D}$ (green surface) separates pitchforks and Hopf bifurcations regimes. Hopf bifurcation type switches at a Bautin bifurcation (blue plane independent of $q$). (A) $J_{11}=1$, $J_{21}=3$; the Bautin and 5-branch pitchfork intersect along a line projected on the $(q,J_{21})$ plane in (A'), where the system may present a codimension-three bifurcation. (B) $J_{11}=1.2$ and $J_{21}=1$; Bautin plane does not intersect the Hopf bifurcation surface within the range of parameters studied, and the type of Hopf bifurcation is always supercritical. (C) Bifurcation surfaces as a function of $q$ and $\mu=tJ_{11}J_{22}$; surface plotted: $\beta_{c}$, $\beta_{p1}$, $\beta_{p2}$, $\beta_{h}$ and the 5-branch pitchfork bifurcation (orange line) for $J_{11}=0.9$ and $J_{22}=0.3$.}
\label{fig:TheoreticalCurves}
\end{figure}

The previous section identified theoretically the locus and type of bifurcations arising at the disordered state: pitchfork bifurcations, Hopf bifurcations, and codimension two Bautin and degenerate pitchfork bifurcations, together with a relatively singular 5-branch pitchfork bifurcation. The possible presence of these bifurcations were parameterized by the noise parameter $\beta$, and further depend on other parameters of the systems, particularly the interaction coefficients $J_{ij}$ and the proportion of hipsters $q$. In Figure~\ref{fig:TheoreticalCurves}, we computed the surfaces associated with the conditions related with the presence of pitchfork bifurcations or Hopf bifurcations (two regions separated by the 5-branch pitchfork bifurcation point) as well as the Bautin bifurcation, as a function of $J_{12}, \, J_{22}$ and $q$. In that figure, the surface $D=0$ (orange surface) separates regimes associated with a 
single pitchfork bifurcation from the parameters associated with two pitchfork bifurcations. The green surface represents the condition associated with the 5-branch pitchfork bifurcations, separating the parameter space into a region where the disordered state loses stability through a pitchfork bifurcation ($J_{22}$ larger than the value associated with that surface) from the region where the instability arises through a Hopf bifurcation, thus associated with periodic solutions (below the surface). The Hopf bifurcation is supercritical if $J_{22}$ is below the blue surface that represents the Bautin bifurcation, corresponding to the change of Hopf bifurcation type. In Fig.~\ref{fig:TheoreticalCurves}(A), the type of Hopf bifurcation thus depends on $J_{22}$, while in Fig.~\ref{fig:TheoreticalCurves}(A), all Hopf bifurcations are supercritical. Interestingly, for the parameters used to compute Fig.~\ref{fig:TheoreticalCurves}(A), the 5-branch pitchfork bifurcation and the Bautin bifurcation surfaces intersect along a line (orange line) that we projected in Fig.~\ref{fig:TheoreticalCurves}(A') on the plane $(J_{12},q)$, and which corresponds to higher-order degeneracy of the system corresponding to a codimension-three bifurcation. 

\subsubsection{Heuristic discussion}
The conditions associated with the stability of the disordered state are much more complex than in the symmetric interaction case. First of all, while no bifurcation was possible in the symmetric system with $q>1/2$, this is no more the case in the asymmetric interaction case. In particular, we observe that the disordered state will lose stability for sufficiently high $\beta$, in the case where $J_{11}$ and $J_{22}$ are not both zero, as soon as:
\[q<\frac{J_{11}}{J_{11}+J_{22}}.\]
Interestingly, this condition is independent of the cross-interaction coefficients $J_{12}$ and $J_{21}$. This condition can be interpreted readily in terms of behaviors of mainstreams and hipsters, particularly clearly in the limit cases where $J_{11}\ll J_{22}$ or $J_{11}\gg J_{22}$. In the former case, the mainstreams interact heavily together, much more strongly than in the interactions within the hipster population. They shall thus rapidly settle on a consensus at high enough $\beta$, and the disordered state cannot be sustained. In that case, regardless of the proportion of hipsters present, an instability of the disordered state occurs. In the latter case, the mainstreams show very little interaction compared to the hipsters, those will maintain a disordered state even at high $\beta$, and any perturbation of this state towards a biased consensus will be rapidly repressed by the hipsters as soon as $q>0$, and thus only at $q=0$ an instability can occur. For intermediate values, the above condition delineates the boundary in the parameter space $q,J_{11}$ and $J_{22}$ between these two regimes. 

When an instability of the disordered state occurs, the type of dynamics tightly depends on the cross-population interaction terms $J_{12}$ and $J_{21}$. Two main situations appear: stabilization on a consensus, or oscillations. When cross-populations interactions are weak compared to the self-interaction coefficients and in the situation where the disordered state is unstable, mainstreams will rapidly find a consensus, to which hipsters will oppose, but this opposition will not be sufficient to have mainstreams change their state, and a consensus emerges, to which all hipsters will have no choice but opposing to, thus all taking the same decision. This corresponds to the pitchfork instabilities. In contrast, when both cross-population interaction coefficients are large, an instability will, again, typically yield a rapid consensus driven by mainstreams, to which hipsters will oppose, and for sufficiently large impact of hipster choices on mainstreams, mainstreams will collectively change state to align to the opposing hipster state, and this switch will periodically repeat, yielding oscillatory responses. Again, in that case, hipsters will all take the same decision and remain synchronized all in the same state at the same time, although this state may evolve. Rigorously, the transition between regimes associated with the emergence of consensus and those yielding oscillations is related to the condition $T>2\sqrt{-D}$, valid whenever $\mu=J_{12}J_{21}$ is large enough. Rigorously, periodic responses arise as $\mu$ exceeds a critical value $\mu_{c}$ at which point the Jacobian determinant (monotonic in $\mu$) exceeds a critical value:
\begin{equation}\label{mucrit}
\mu_{c}=\frac{1}{\beta_{c}^{2}q(1-q)}+ J_{11}J_{22}.
\end{equation}

This is visible in Fig.~\ref{fig:TheoreticalCurves} and the prominence of the Hopf instability at the expense of the pitchforks as $\mu$ increases.

\subsubsection{Pitchforks and bifurcations of the non-disordered states. }
\begin{figure}[h]
\begin{center}
\includegraphics[width=0.8\textwidth]{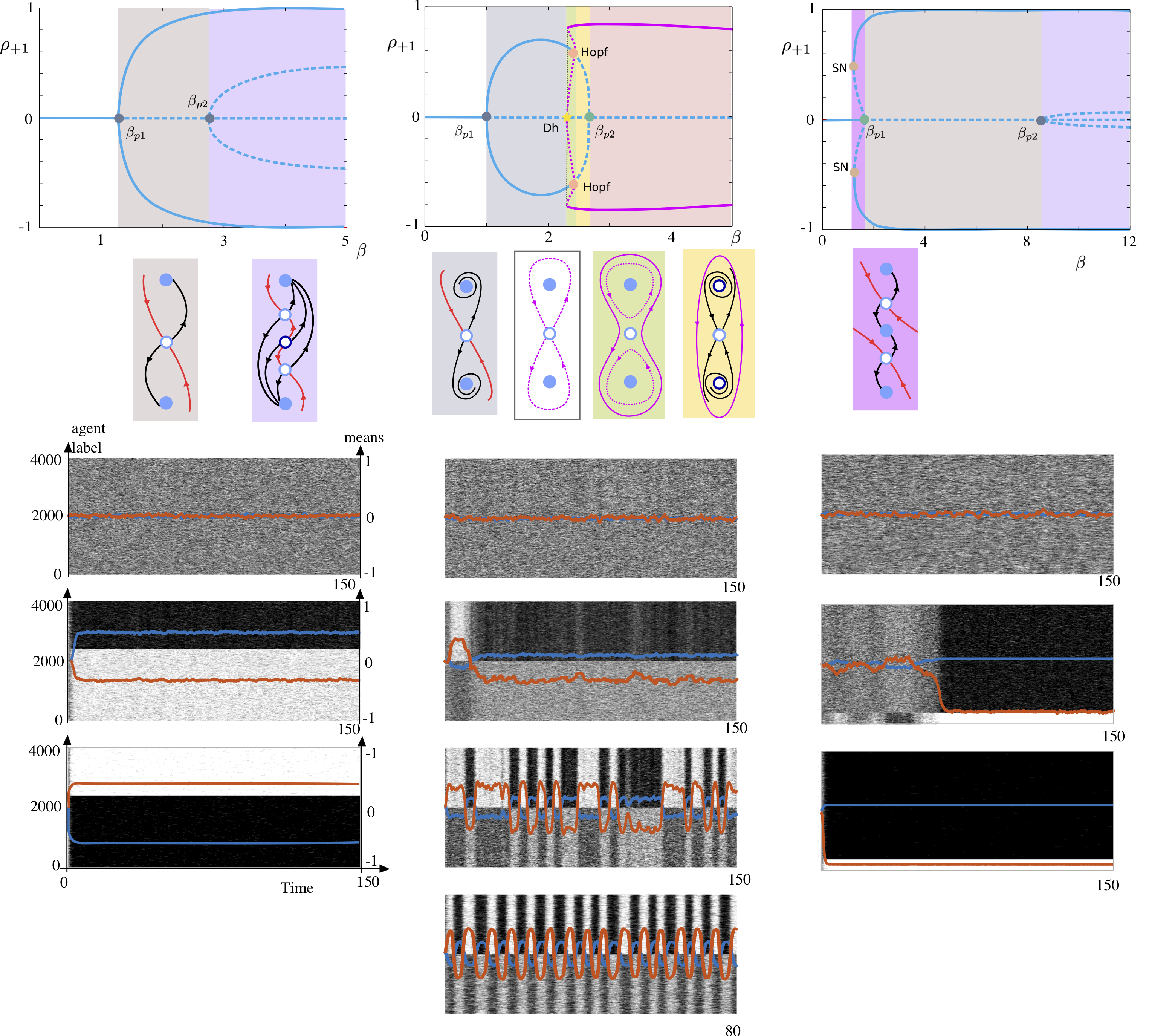}
\end{center}
\caption{Pitchfork bifurcations with $D<0$ and $T>2\sqrt{-D}$. (A) $q=0.6$, $J11=3.25$, $J_{12}=1$, $J_{21}=2$ and $J_{22}=0.25$. Both bifurcations are supercritical (respectively, $\gamma\simeq -0.54$ and $\gamma\simeq-0.38$). Blue solid lines: stable fixed points, dashed lines: unstable fixed points. (B) $q=0.5$, $J_{11}=3$, $J_{12}=4.5$, $J_{21}=0.5$ and $J_{22}=0.25$; pitchfork at $\beta_{p1}$ is supercritical ($\gamma\simeq -8.91$) and at $\beta_{p2}$ subcritical ($\gamma\simeq 0.66$): the two branches of non-disordered equilibria collide, and a subcritical Hopf bifurcation arises (dashed pink line) undergoing a fold-double-homoclinic bifurcation (Dh), associated with the presence of a stable periodic orbit (pink solid line) persisting for larger values of $\beta$ tested. (C) $q=0.1$, $J_{11}=1.2$, $J_{12}=1$, $J_{21}=5$, $J_{22}=3.5$. The pitchfork at $\beta_{p1}$ is subcritical ($\gamma\simeq 0.08$), and at $\beta_{p2}$ supercritical ($\gamma\simeq 1.56$). A saddle-node bifurcation is numerically identified on the branch of the unstable fixed points emerging from $\beta_{p1}$. Phase planes are hand-drawn below the diagrams in 7 typical situations identified (matching the colors on the diagrams), and responses of a stochastic network with $n=4\,000$ agents confirm these dynamics for values of $\beta$ in each regime: (A): $\beta=0.5,\,2,\,4$, (B): $\beta=0.5,\, 1.5,\, 2.3,\, 3$, (C): $\beta= 1,\,1.57, 4$. Type of each individual depicted in color (+1: white, -1: black), mainstream are on top, and average type for hipsters (blue) or mainstreams (red) are added on top of this diagram. 
Bifurcation diagrams generated with XPP Aut~\cite{ermentrout2002simulating}, simulations performed on custom code on Matlab. }
\label{fig:pitchforks}
\end{figure}

The bifurcations identified reveal the emergence of up to 4 non-disordered equilibria, with various stability properties, associated the pitchfork bifurcations. The closed form condition for the type of pitchfork bifurcation derived in proposition~\ref{prop:pitch} (the value of $\gamma$) was used to identify parameters associated with the various types of bifurcations possible. To this purpose, we randomly sampled the parameter space ($10^{7}$ points) and computed the values of $\gamma$ for the two bifurcation point. In Figure~\ref{fig:pitchforks}, we represent three typical cases: the case where both pitchfork bifurcations $(\beta_{p1},\beta_{p2})$ are supercritical, or when one is supercritical and the other subcritical. Despite extensive parameter space exploration, no situation with both pitchforks being subcritical was identified. 

Our numerical simulations confirm rigorously the predicted type of pitchfork bifurcation. 

Moreover, in these parameter regimes we computed the bifurcation diagram of the system to identify the possible presence of bifurcations at the non-disordered states. In the case where the two pitchfork bifurcations are supercritical (Fig~\ref{fig:pitchforks}A), no additional bifurcation is observed. The system features up to five equilibria in total and up to two stable fixed points. The unstable equilibria organize the attraction basins of the stable fixed points, particularly the stable manifolds of the saddles. Typical phase portraits in the presence of non-trivial fixed points show the general organization of those stable manifolds in both cases. Moreover, we simulated a stochastic hipster network for the same parameters, and found a very good consistency with the behaviors predicted by the mean-field limit. 

In the case where $\beta_{p1}$ is supercritical and $\beta_{p2}$ subcritical (Fig.~\ref{fig:pitchforks}B), the branch of stable fixed points emerging from $\beta_{p1}$ collapses with the branch of unstable fixed point associated with the pitchfork bifurcation at $\beta_{p2}$. A Hopf bifurcation on the non-trivial equilibria arises, and embodies the transition between stable and unstable fixed points. Numerically, we found that this Hopf bifurcation is subcritical, and thus associated with the emergence of unstable periodic orbits remaining either in the positive or in the negative trend half-planes. The amplitude of these orbits grow away from the Hopf bifurcation point until hitting the trivial equilibrium on a double-homoclinic bifurcation. At this point also, the system undergoes a fold of limit cycles, and a branch of stable orbits switching between positive and negative trends and enclosing the fixed points (green and pink regions of the diagram). The dynamics of the system for these parameters is much more complex: in particular, we observe the emergence of periodic trajectories, as well as tri-stability between consensus equilibria and a periodic orbit. Finite size effects in the stochastic system within this regime displays random transitions between those three attractors. As indicated in the bifurcation diagram depicted in Supplementary Fig.~\ref{fig:codim2}(B), the branch of subcritical Hopf bifurcation exists between the point $P$ at which the pitchfork bifurcation at $\beta_{p2}$ switches from sub- to super-critical, at which point emerges a saddle-node bifurcation. The subcritical Hopf bifurcation and this saddle node bifurcation collapse at a Bodgdanov-Takens bifurcation, beyond which the Hopf bifurcation disappears. 

Eventually, the case where $\beta_{p1}$ is subcritical and $\beta_{p2}$ super critical is similar in many ways to the case where both pitchforks are subcritical (Fig.~\ref{fig:pitchforks}C). The main notable distinction is the presence of three stable fixed points in the vicinity of $\beta_{p1}$: the disordered equilibrium and two consensus, separated by two saddles whose stable manifolds organize the attraction bassins. Finite-sized stochastic networks may thus switch between those attractors, as we display in the evolution below the diagram (second panel).

\subsubsection{Hopf bifurcations and noise-induced synchronization}
Our theoretical analysis also described the presence of Hopf bifurcations, directly associated with the presence of oscillations in the vicinity of the instability, whose stability was characterized. We numerically computed the bifurcation diagrams associated with these parameter sets. Contrasting with the delay-induced synchronization, we found here systematically the emergence of a pair of stable non-disordered equilibria at high $\beta$, connected with the cycles through a homoclinic orbit. 

\begin{figure}
\begin{center}
\includegraphics[width=.8\textwidth]{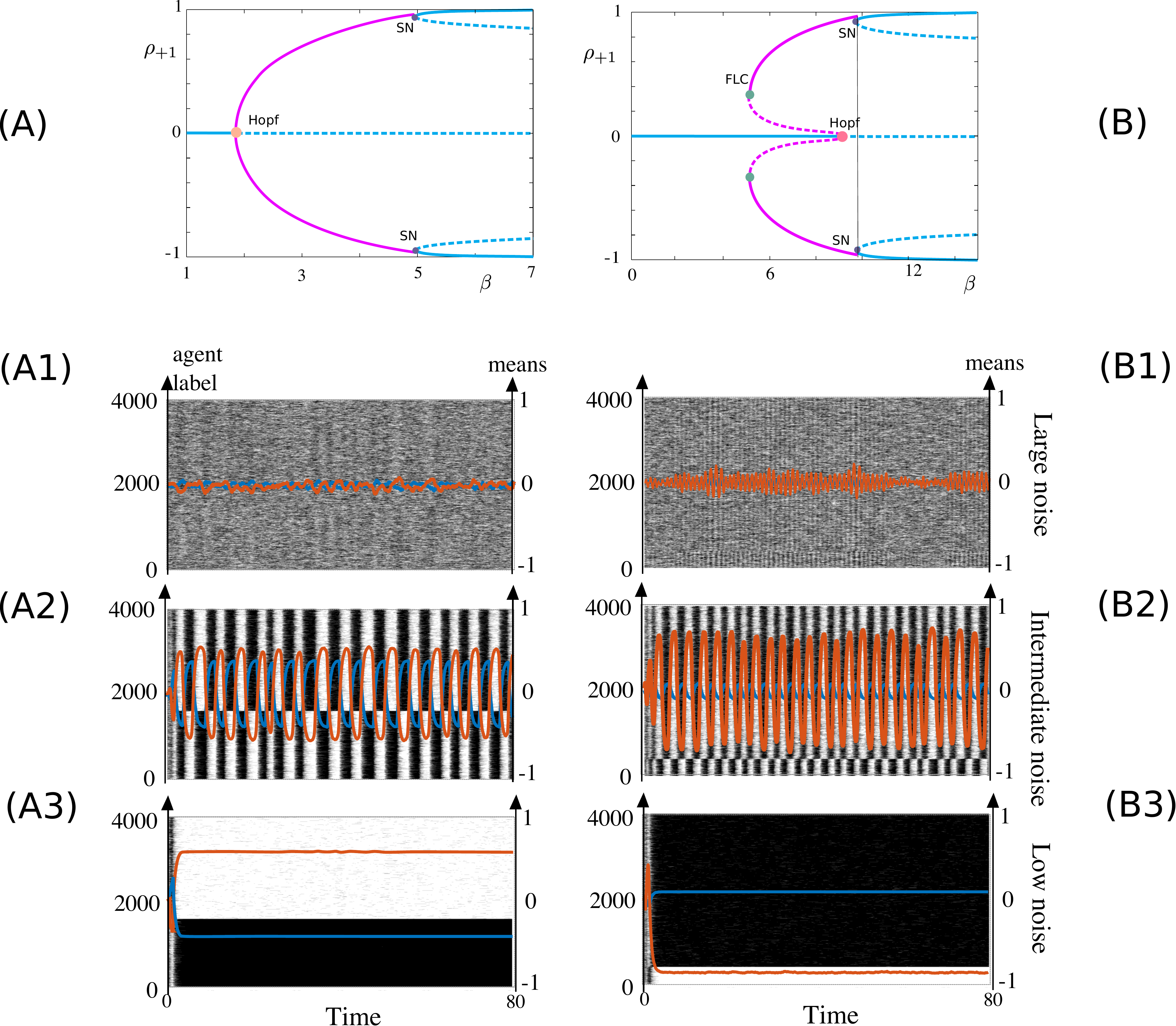}
\end{center}
\caption{Super- and Sub-critical Hopf bifurcations in the asymmetric Hipster model. (A, subpercritical) $q=0.4$, $J11=3.1$, $J_{12}=3.5$, $J_{21}=3$ and $J_{22}=2$. (B, subcritical) $q=0.1$, $J11=0.8$, $J_{12}=5$, $J_{21}=3$ and $J_{22}=5$. Top: bifurcation diagrams. Bottom panels: simulations of the network model, with $n=4\,000$, and various values of the noise parameter $\beta$: A1: $\beta=1$, A2: $\beta=4$, A3: $\beta=5.5$, B1: $\beta=3.5$, B2: $\beta=6$, B3: $\beta=11$. }
\label{fig:Hopfs}
\end{figure}

From the network viewpoint, this transition is surprising and somewhat paradoxical. Indeed, this transition indicates the emergence of a structured rhythmic activity triggered by an increase in the randomness of the decisions of the agents. A similar transition was observed in neural networks~\cite{touboul2012noise} or abstract models~\cite{scheutzow1985some,scheutzow1985noise}. This is a somewhat mysterious transition. Indeed, while, on one hand, it is clear that high levels of noise (low $\beta$) lead to the absence of stable consensus equilibrium because the randomness in the transition dominates, and on the other hand, at very low noise levels (i.e., high $\beta$), the mainstream agents can find a consensus, imposing hipsters to oppose to it, it remains unclear how noise can trigger the emergence of periodic responses. We suggest that noise triggers switches between the two symmetric consensus states that may arise; noise facilitates transitions of mainstreams and hipsters to opposite regimes, and these marginal transitions may amplify under the condition of the presence of Hopf bifurcations, i.e. when $T^{2}<-4D$, or:
\[((1-q)J_{11}+qJ_{22})^{2}<4q(1-q)J_{12}J_{21},\]
which provides a weighted version of a condition indicating that cross-population interactions dominate intra-population interactions. 

We eventually note that sub- or super-critical Hopf bifurcation are associated with clearly distinct behaviors of the networks at low $\beta$: supercritical regimes show a progressive build-up of synchronization with a disappearance of the disordered regime, while in the sub-critical Hopf bifurcation regime, a sudden highly synchronized regime emerges and is stable, while the disordered state conserves stability. 

\begin{figure}
\begin{center}
\includegraphics[width=.7\textwidth]{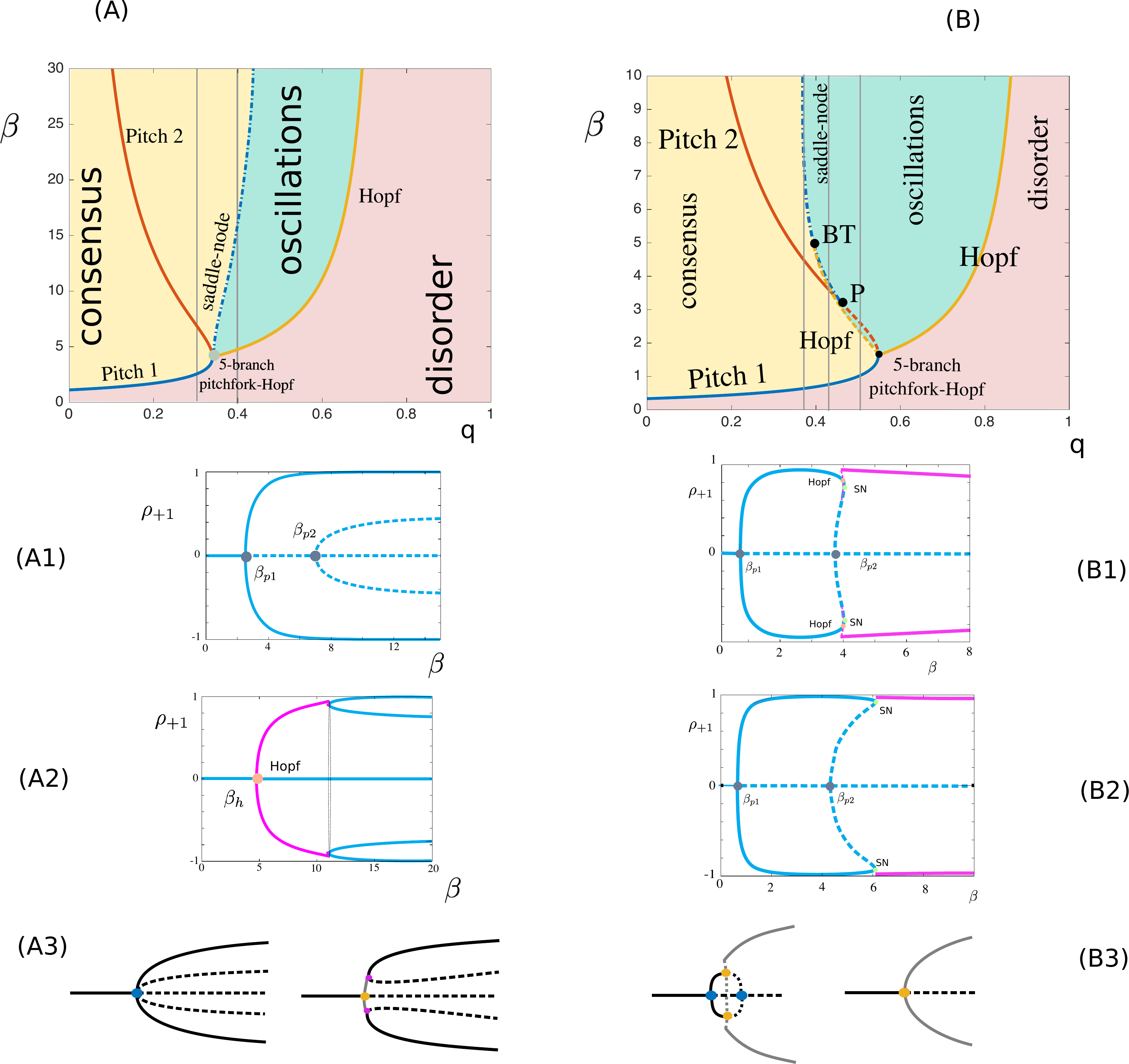}
\end{center}
\caption{Codimension-two diagrams. Top: Two codimension two diagrams, (A) $J_{11}=0.9$, $J_{12}=0.9$, $J_{21}=0.6$ and $J_{22}=0.3$, where the pitchfork bifurcation $\beta_{p2}$ is always supercritical, or (B) parameters associated with Fig.~\ref{fig:pitchforks}, center. Blue: $\beta_{p1}$, Red: $\beta_{p2}$ (solid: supercritical, dashed: subcritical), yellow: $\beta_{h}$ depicted analytically. Dotted blue line is the saddle-node bifurcation of non-disordered equilibria and dotted yellow line is the subcritical Hopf bifurcation on non-disordered states, both computed using Matlab Matcont package~\cite{dhooge2006matcont,dhooge2003matcont}. BT: Bogdanov-Takens, P: transition between sub- and super-critical pitchfork bifurcations.
(A1) $q=0.3$, (A2): $q=0.4$ (grey lines in (A)). (A3): idealized unfolding of the 5-branch pitchfork bifurcation (black: fixed points, gray: cycles, circle color as in (A-B), except saddle-node represented in pink). (B1) $q=0.42$, (B2): $q=0.38$, and the case $q=0.5$ is Fig.~\ref{fig:pitchforks}. (B3) idealized unfolding of the 5-branch pitchfork. }
\label{fig:codim2}
\end{figure}

\subsubsection{Codimension-two bifurcations}
We summarize these results in the codimension-two bifurcation diagrams of Fig.~\ref{fig:codim2}. In that figure, we depict, as a function of $q$ and $\beta$, the bifurcation lines associated with the local bifurcations at the disordered state, as well as the saddle-node and possible Hopf bifurcation associated with consensus equilibria. These diagrams allows appreciating how the bifurcation lines, and particularly the 5-branch pichfork bifurcation, organize the dynamics. A clear distinction in the emergence of oscillations is observed in this figure: in the case where $\beta_{p2}$ is supercritical at the 5-branch pitchfork bifurcation point (Fig.~\ref{fig:codim2}), we observed that the graph of the saddle-node bifurcation is an increasing function in the plane $(\beta,p)$, indicating the presence of noise-induced oscillations in the system for $q$ larger than the value associated with the 5-branch pitchfork bifurcation. In contrast, in the case where $\beta_{p2}$ is subcritical at the 5-branch pitchfork bifurcation, the oscillations are sustained and no noise-induced bifurcation occurs for $q$ larger than the bifurcation point; the saddle-node bifurcation forms a decreasing graph in the plane $(\beta,p)$, and no noise-induced oscillation emerge for $q$ larger than the value associated with the 5-branch pitchfork bifurcation, but such noise-induced phenomena arise for $q$ smaller, and persist even for small noise. 

\subsubsection{Asymmetric interactions and effective delays} 
While the framework used in the asymmetric interaction case differs from direct delays in the communication, we argue that these asymmetries may cause synchronization due to the emergence of an effective feedback delay of the decisions of individuals, not simply due to transmission or reaction to perceived trends, but rather due to the intrinsic time taken by individuals to respond to changes in the dynamics. In that sense, one could expect that differences in the timescales of the response of individuals could cause similar phenomena. There are at least two manners to take into account the timescales of reaction of individuals in non-delayed systems. First, the reactivity of individuals, namely the speed at which they aim at either decreasing or increasing their dissimilarity to others, is of importance. For instance, one could imagine that hipsters are more sensitive to imbalances that mainstreams. The sensitivity to an imbalance is controlled by the parameter $\beta$ that also controls the noise in the decisions. If hipsters and mainstream adjust their style with distinct sensitivities, but the same maximal rate, one could consider an asymmetric model with two sensitivities $J_{11}=J_{12}=\beta_{+1}$ and $J_{11}=J_{12}=\beta_{-1}$ (and fix $\beta=1$ without loss of generality), with $\beta_{\pm 1}$ being the parameter associated with populations $\pm 1$. That case thus falls into the degenerate case $D=0$, in which case the only bifurcation arising is a pitchfork bifurcation (see proposition~\ref{prop:pitch}). An heterogeneous sensitivity will thus not create oscillations. 

\begin{figure}
\begin{center}
\includegraphics[width=.7\textwidth]{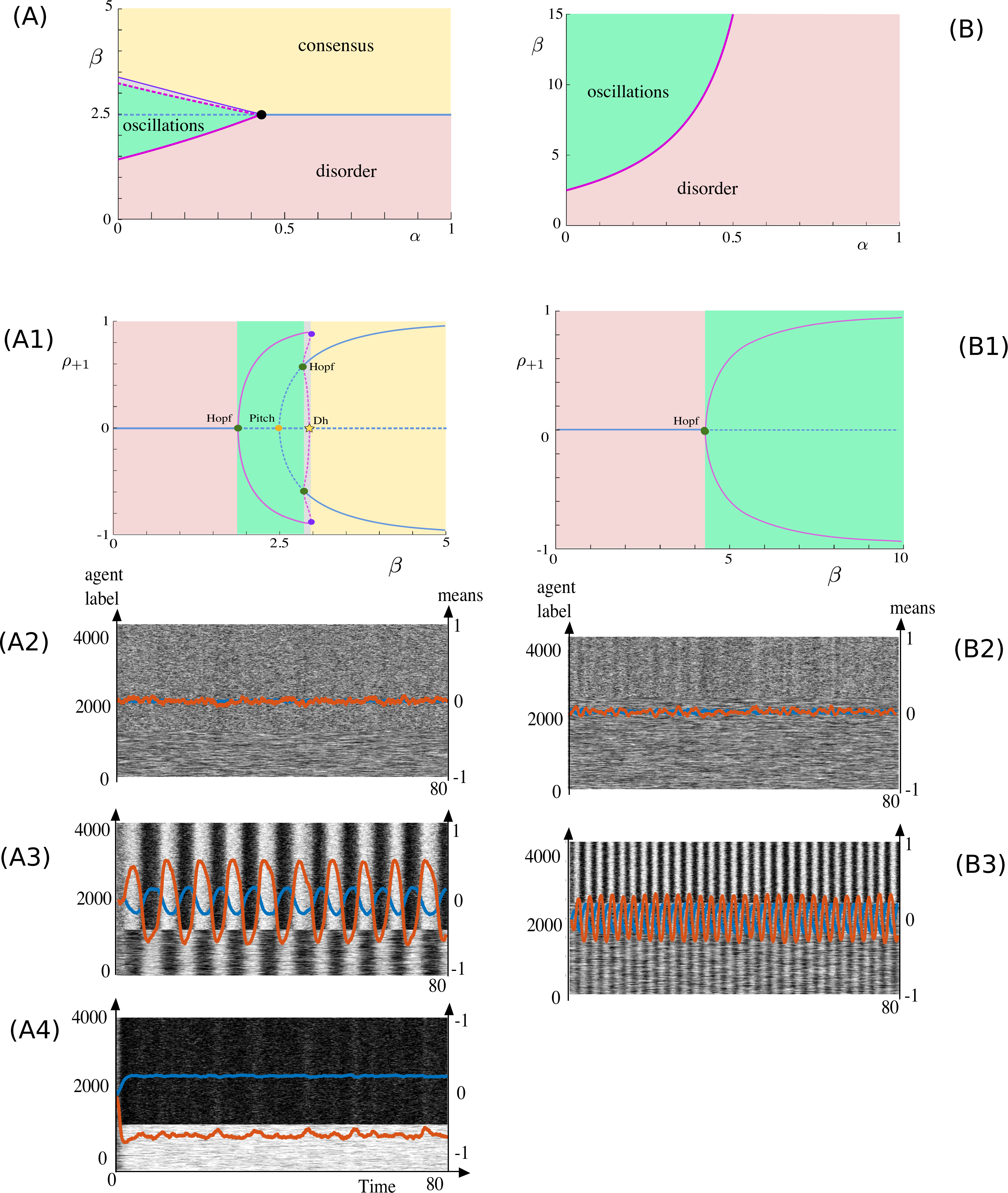}
\end{center}
\caption{Role of the timescales in synchronization: a symmetrically interacting hipster model with distinct transition rates function: $\f_{1}(x)=2(1+\tanh(x))$ for mainstreams, and $\f_{2}=\alpha \f_{1}$ for hipsters. Top: codimension two bifurcation diagrams as a function of $\alpha$ and $\beta$ for (A) $q=0.3$, (B) $q=0.6$. Pink lines: Hopf bifurcations (solid: supercritical, dashed: subcritical), blue: supercritical pitchfork bifurcation (dashed: yielding unstable fixed points, solid: yielding stable fixed points) (computed with Matcont). Purple: fold of limit cycles, hand-drawn. (A1-B1) codimension-one diagram for $\alpha=0.2$: blue lines are fixed points, pink lines cycles, solid / dashes: stability. (A1): the disordered state loses stability at a Hopf bifurcation, then undergoing a pitchfork bifurcation yielding consensus equilibria. These equilibria gain stability through a Hopf bifurcation. Cycles collide at a double-homoclinic fold of cycle bifurcation, similar to the one observed in Fig.~\ref{fig:pitchforks}, middle. (B1): the disordered state loses stability in favor of a cycle that persists for larger $\beta$. (A2-A4) network simulations with $n=4\,000$ and $\beta=1,\, 2.5$ or $3.5$ respectively. (B2-B3) $\beta= 3$ or $6$. }
\label{fig:DifferentTimescales}
\end{figure}

However, a more relevant way to model the slowness in taking decisions consists in scaling the rate at which mainstreams and hipsters switch states. In the above study, we considered that all agents have a rate modeled by a single map $\f$, and the distinction were only in the argument of that function. To take into account the specific timescales of both populations and test their impact on the solutions, we shall thus consider mainstream transitions with a rate $\f_{1}(x)=\f(x)$ and, for hipsters, $\f_{-1}(x)=\alpha \f_{1}(x)$. Values of $\alpha$ greater than 1 correspond to situations where hipsters are faster than mainstreams, and $\alpha<1$ to hipsters slower than mainstreams. It is not hard to generalize the study above to take into account distinct timescales. We analyzed this system and indeed found, consistently to the asymmetric interaction case, the presence of noise-induced synchronizations when $\alpha<1$, i.e. when hipsters are slower than mainstreams in their reaction to trends. In Fig.~\ref{fig:DifferentTimescales}, we quantified this effect computing the bifurcation diagram of the limit system for a symmetric interaction system as a function of the typical reactivity rate of hipsters compared to mainstreams $\alpha$. We observed that a transition to synchronized oscillations occurs both in the case where hipsters are a minority ($q=0.3$, left) or a majority ($q=0.6$, right). In the case where hipsters are the minority, a pitchfork also arises, as observed in the case $\alpha=1$, and the oscillations stop at a homoclinic bifurcation, indicating again the presence of noise-induce oscillations. This transition is absent in the case $q>1/2$. Heuristically, this transition relies on similar phenomena as described in a purely delayed situation: when hipsters rate of change is too low, i.e. hipsters are too slow, they will leave room for a transient synchronization of mainstreams, to which they will oppose and, depending upon parameters, may revert, leading to reiterate the process periodically. 

\section{Conclusion}
In this paper, we introduced and studied a model of interacting agents with two classes: mainstreams, that follow the majority, and hipsters, that aim at opposing to it. We showed that, in contrast to cooperative systems, populations of individuals that take decision in opposition to the majority undergo phase transitions to oscillatory synchronized states. These oscillations may emerge when taking into account delays, either modeling the time it takes for each individual to react to perceived trends (section~\ref{sec:delays}) or when delays are heterogeneous and emerge due to the transmission of information in spatially extended systems (section~\ref{sec:binary_ring}). Similar synchronization phenomena arise when the interactions between mainstreams and hipsters are not identical (section~\ref{sec:asym}), and in that case complex transitions occur that tightly depend on the relationship between inter- and intra-population interaction coefficients.

The analysis of the relatively simple binary hipster model allowed going quite far in the understanding of the concurrent role of noise, delays, proportions of hipsters and mainstream individuals and relative impact of inter- and intra-population interaction in the emergence of synchronization. Overall, we observed that it remains in all cases a difficult task for the hipsters to avoid synchronization and keep opposing to the majority in a consistent manner, and tightly relies on all parameters. Along the way, we uncovered several points that are well worth studying in depth. For instance, the behavior of a system with an equal proportion of hipsters and mainstreams appears to be a singular phase transition in which the whole population tends to randomly switch between different trends, and would be very interesting to further characterize. 

\begin{figure}
\includegraphics[width=.7\textwidth]{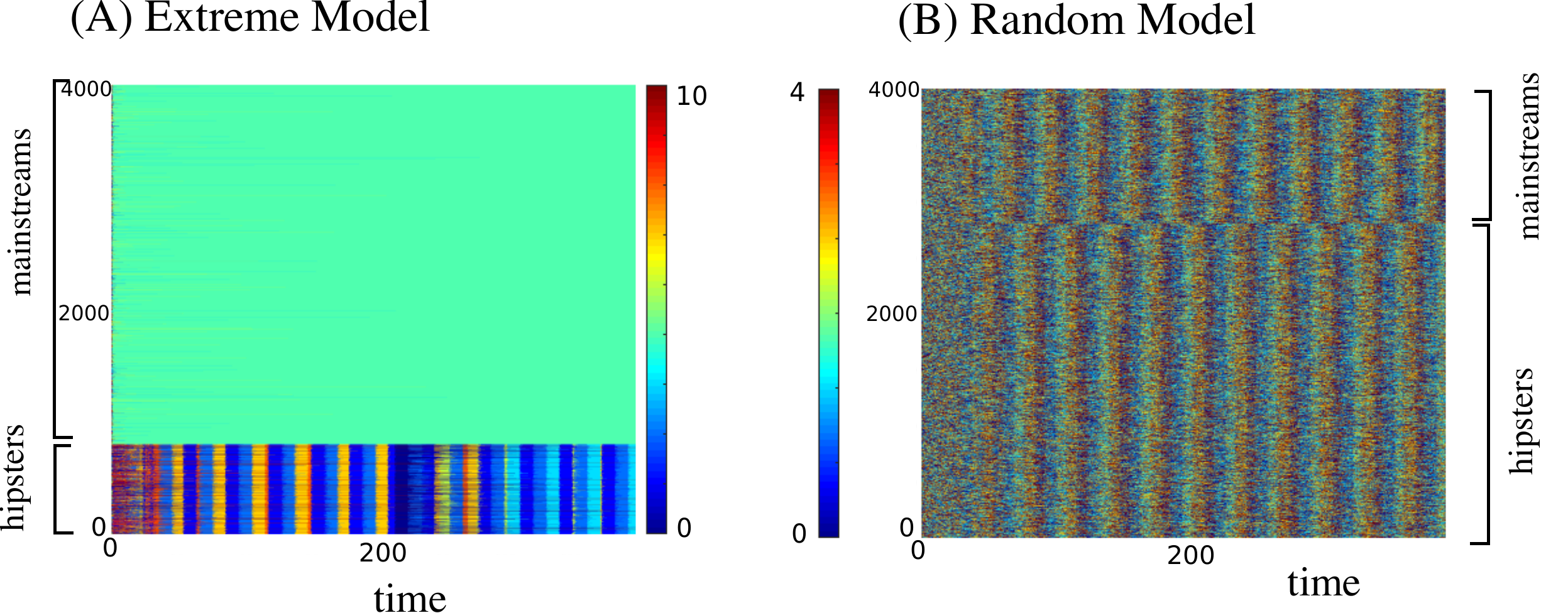}
\caption{Synchronization in models with $P>2$ choices. (A) Extreme scenario: jump occur, similar to the binary model, according to conformity to the trend, and switches depend on occupation levels: hipsters (mainstreams) switch to the least (most) occupied state. $P=10$ choices, $\beta=2$, $q=0.2$ and fixed delay $\tau=10$. (B) Random scenario: Pott's model with mainstreams and hipsters and delay: when an individual switches state, it chooses uniformly at random among other states. $P=4$, $\beta=15$, $q=0.7$ and $\tau=15$. }
\label{fig:Pchoices}
\end{figure}

In a sense, one may believe that the oscillation observed may be an artifact of the excessive simplicity of the model only considering binary choices. Indeed, this simplification may naturally lead to synchronization because of the absence of sufficient alternatives. For instance, coming back to the case of hipsters, if a majority of individuals shave their beard, then most hipsters will want to grow a beard, and if this trend propagates to a majority of the population, it will lead to new, synchronized, switch to shaving. But what if one can grow a mustache, a square beard or a goatee, would that diversity of choices allow hipsters to be as different as they can? In other words, would the hipster effect synchronizing all anticonformist dissolve in a complex world? We will study in depth this question in a forthcoming paper. However, we already present, in Fig.~\ref{fig:Pchoices}, cases where individuals have more than two possible states and react with delays, and observe, consistently with the binary model, synchronization of slow hipsters, even when they belong to the minority of the population. 

This study and simple models introduced thus opens the way to a better understanding of synchronization and correlations in statistical models. Simple systems have proved invaluable for getting insight into more complex systems and may provide access to universal behaviors governing complex models developed in applied domains, for instance in sociophysics, or financial applications whereby speculators may make profit when taking decisions in opposition to the majority in stock exchange~\cite{challet2013minority,galam2012sociophysics}. 

\bigskip 

\appendix
\centerline{\Large \textsc{Appendix}}
\section{Brief discussion on the impact of delays in the asymmetric binary hipsters model}
In section~\ref{sec:asym}, we studied the hipster model with asymmetric interactions, and showed that delays are not necessary to generate oscillations in the hipster model, provided that the interaction coefficients satisfy specific relationships (see also~\cite{collet2016rhythmic}). Adding delays to this system may be much more challenging in the case of multiple or distributed delays that depend on the populations considered. The problem however largely simplifies when considering a single delay (an identical delay in the communication between all types of individuals). A related system was studied in~\cite{faria:00} in the context of neural networks with memory, in a distinct system but with closely related linearization. Similar development can be performed in the model at hand, and we outline the methodology below. 

\begin{figure}
\begin{center}
	\includegraphics[width=\textwidth]{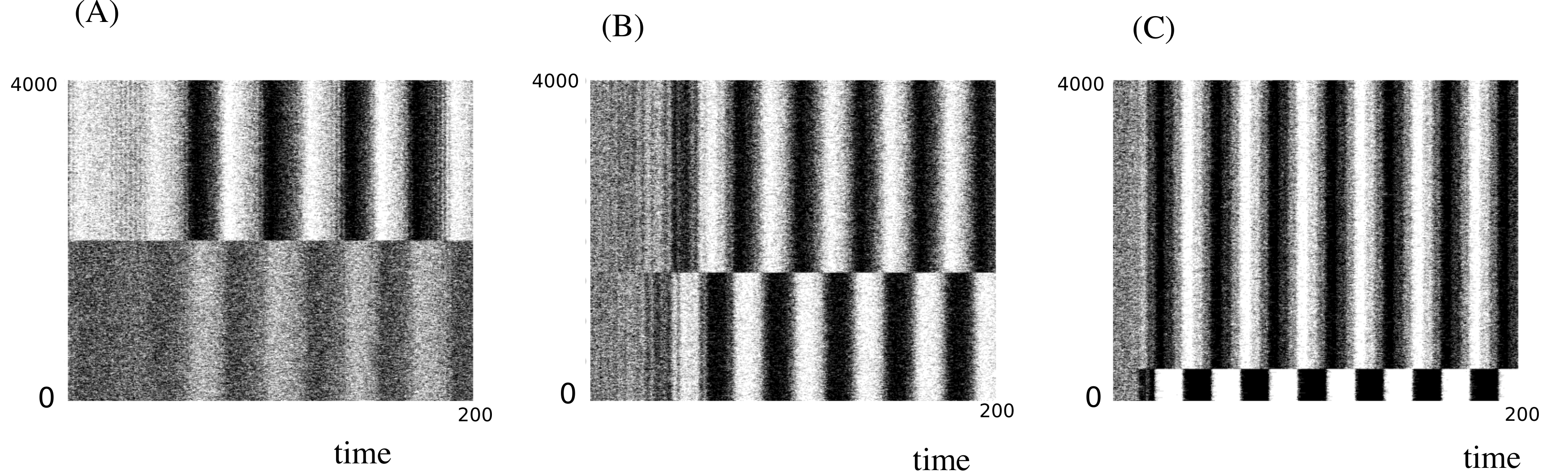}
\end{center}
\caption{Delays advance the emergence of oscillations in the hipster model with asymmetric interactions. Parameters as in (A) Fig.~\ref{fig:pitchforks}(B) with $\beta=2$ and $\tau=10$, (B) Fig.~\ref{fig:Hopfs} (A) with $\beta=1.5$ and $\tau=4$ and (C) Fig.~\ref{fig:Hopfs} (B) with $\beta=3.5$ and $\tau=4$. In all cases, the parameters correspond to the absence of oscillations in the non-delayed system, and delays induce synchrony.}
\label{fig:Asym_Delay}
\end{figure}

We assume in this section that $p_{\eps,\eps'}(j,\tau)=\delta_{J_{\eps,\eps'}}(j)\delta_{\tau_{0}}(\tau)$. The Kolmogorov equations for the large $n$ limit~\eqref{eq:Kolmogorov} now read:
\begin{equation}\label{eq:AsymDelay}
\begin{cases}
\dot{\rho}_{+1}&=-2 \left[\rho_{+1}(t)+\tanh\left(-\beta \left(J_{11} (1-q) \rho_{+1}(t-\tau_{0})+J_{12} q \rho_{-1}(t-\tau_{0})\right)\right)\right]\\
\dot{\rho}_{-1}&=-2 \left[\rho_{-1}(t)+\tanh\left(\beta \left(J_{21} (1-q) \rho_{+1}(t-\tau_{0})+J_{22} q \rho_{-1}(t-\tau_{0})\right)\right)\right]
\end{cases}
\end{equation}
Pitchfork bifurcations are identical to the ones found in the absence of delay, but delays can indeed induce Hopf bifurcations not present in the instantaneous communication case. To identify these points, we derive the dispersion relationship and, looking for solutions of the type $(h_{1},h_{2})e^{\zeta t}$, we obtain the system:
\begin{equation}
\begin{cases}
\zeta h_{1}= (-2+2\beta J_{11} (1-q) e^{-\zeta \tau_{0}})h_{1}+ 2\beta J_{12} q\,e^{-\zeta \tau_{0}}h_{2}\\
\zeta h_{2}= (-2\beta J_{21} (1-q)e^{-\zeta \tau_{0}})h_{1} -(2+2 \beta J_{22} q e^{-\zeta \tau_{0}})h_{2}.
\end{cases}
\end{equation}
This linear system has non-trivial solutions when $\zeta$ solves the dispersion relationship found as the determinant of the above system:
\[(-2-\zeta+2\beta g_{11} e^{-\zeta \tau_{0}})(-2-\zeta-2\beta g_{22} e^{-\zeta \tau_{0}}) + 4 \beta^{2}g_{12}g_{21} e^{-2\zeta \tau}=0,\]
which we can rewrite as the determinant equation in the absence of delay: 
\[1-\xi T -\xi^{2}D=0\]
for $\xi=2\beta (2+\zeta)^{-1}e^{-\zeta\tau_{0}}$. In the absence of delay, we have been interested in the real and positive solutions of that equation to find pitchfork bifurcation. Delays provide an interpretation for complex solutions, and constrain the value of the delay. Curves of Hopf bifurcations can thus be found similarly and depend on the level of delay. In Fig.~\ref{fig:Asym_Delay} we show a few examples of trajectories of the system for parameters considered in section~\ref{sec:asym}. We generally observe that delays yield oscillations for values of $\beta$ smaller than in the absence of delay. 

\section{Supplementary Figures}
\newpage
\begin{figure}
\begin{center}
\includegraphics[width=\textwidth]{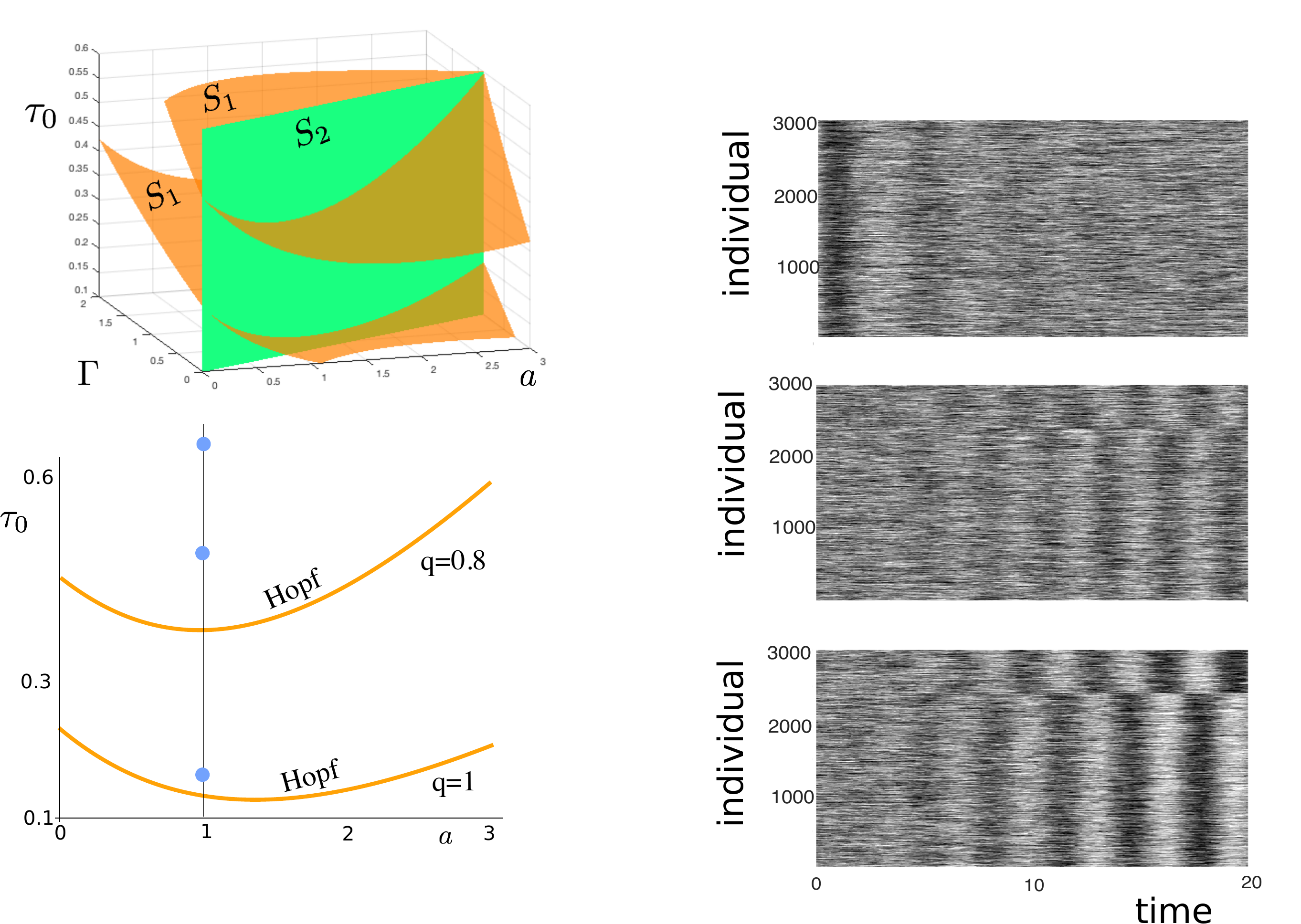}
\caption{Dependence of the Hopf bifurcation curve upon variation of $q$ in the spatially extended case (section~\ref{sec:binary_ring}). Decreasing $q$ tends to stabilize fixed points, and larger delays are necessary to synchronize the system. Upper-left diagram: orange surfaces correspond to $S_{1}$, for $q=1$ (lower surface) or $q=0.8$, and green surface is $S_{2}$, independent of $q$. The intersection of these curves provide the locus of Hopf bifurcations depicted below, and we note on the right the absence of oscillations at $q=0.8$ where a system with $q=1$ oscillates (top), and larger delays reveal those oscillations. }
\label{fig:Space_q}
\end{center}
\end{figure}

\begin{figure}[h]
\begin{center}
\includegraphics[width=\textwidth]{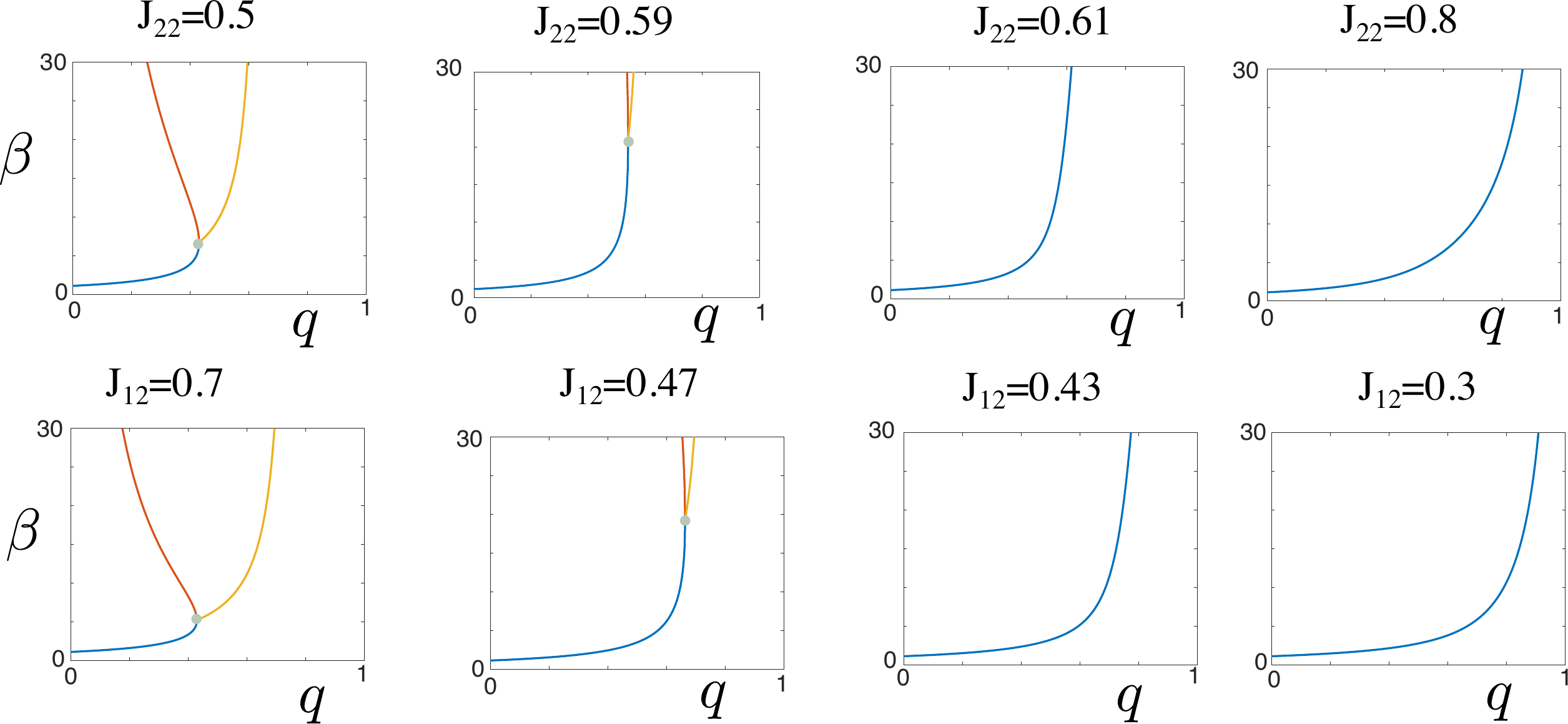}
\caption{Modifications of the codimension-two bifurcation diagram (analytical curves) for parameters associated with Fig.~\ref{fig:codim2} (A) for various values of $J_{22}$ or $J_{12}$. We observe in both cases the disappearance of Hopf bifurcations either by increasing $J_{22}$ or decreasing $J_{12}$, leaving the system with a unique pitchfork bifurcation. In all diagrams, the abscissa is $q$ and the ordinate is $\beta$; blue curve is $\beta_{p1}$ or $\beta_{c}$, red: $\beta_{p2}$, yellow: $\beta_{h}$. }
\label{fig:OtherDiagrams}
\end{center}
\end{figure}

%\begin{figure}
%\begin{center}
%\includegraphics[width=\textwidth]{PitchesHopfs}
%\label{fig:pitch2}
%\end{center}
%\end{figure}
%

\newpage
\bibliographystyle{plain}
\bibliography{hipsters}

\begin{thebibliography}{10}

\bibitem{boltzmann}
L~Boltzmann.
\newblock {\em Lectures on gas theory}.
\newblock Dover, New York, 1895.

\bibitem{brunel-hakim:99}
N.~Brunel and V.~Hakim.
\newblock Fast global oscillations in networks of integrate-and-fire neurons
  with low firing rates.
\newblock {\em Neural Computation}, 11:1621--1671, 1999.

\bibitem{carr}
Jack Carr.
\newblock {\em Applications of centre manifold theory}.
\newblock Springer-Verlag, 1981.

\bibitem{challet2000modeling}
Damien Challet, Matteo Marsili, and Yi-Cheng Zhang.
\newblock Modeling market mechanism with minority game.
\newblock {\em Physica A: Statistical Mechanics and its Applications},
  276(1):284--315, 2000.

\bibitem{challet2013minority}
Damien Challet, Matteo Marsili, and Yi-Cheng Zhang.
\newblock Minority games: interacting agents in financial markets.
\newblock {\em OUP Catalogue}, 2013.

\bibitem{clauset2009power}
Aaron Clauset, Cosma~Rohilla Shalizi, and Mark~EJ Newman.
\newblock Power-law distributions in empirical data.
\newblock {\em SIAM review}, 51(4):661--703, 2009.

\bibitem{collet2016rhythmic}
Francesca Collet, Marco Formentin, and Daniele Tovazzi.
\newblock Rhythmic behavior in a two-population mean-field ising model.
\newblock {\em Physical Review E}, 94(4):042139, 2016.

\bibitem{crisanti-sompolinsky:87b}
A.~Crisanti and H.~Sompolinsky.
\newblock {Dynamics of spin systems with randomly asymmetric bonds: Langevin
  dynamics and a spherical model}.
\newblock {\em Physical Review A}, 36(10):4922--4939, 1987.

\bibitem{crisanti-sompolinsky:87}
A.~Crisanti and H.~Sompolinsky.
\newblock {Dynamics of spin systems with randomly asymmetric bounds: Ising
  spins and glauber dynamics}.
\newblock {\em Phys. Review A}, 37(12):4865, 1987.

\bibitem{dai2013curie}
Paolo Dai~Pra, Markus Fischer, and Daniele Regoli.
\newblock A curie-weiss model with dissipation.
\newblock {\em Journal of Statistical Physics}, 152(1):37--53, 2013.

\bibitem{dairhythmic}
Paolo Dai~Pra, Elena Sartori, and Marco Tolotti.
\newblock Rhytmic behavior in large scale systems: a model related to
  mean-field games.
\newblock {\em preprint}, http://www.math.unipd.it/~daipra/draftGame-Paolo.pdf,
  2017.

\bibitem{dhooge2006matcont}
A~Dhooge, W~Govaerts, Yu~A Kuznetsov, W~Mestrom, AM~Riet, and B~Sautois.
\newblock Matcont and cl matcont: Continuation toolboxes in matlab.
\newblock {\em Universiteit Gent, Belgium and Utrecht University, The
  Netherlands}, 2006.

\bibitem{dhooge2003matcont}
Annick Dhooge, Willy Govaerts, and Yu~A Kuznetsov.
\newblock Matcont: a matlab package for numerical bifurcation analysis of odes.
\newblock {\em ACM Transactions on Mathematical Software (TOMS)},
  29(2):141--164, 2003.

\bibitem{ermentrout2002simulating}
Bard Ermentrout.
\newblock {\em Simulating, analyzing, and animating dynamical systems: a guide
  to XPPAUT for researchers and students}, volume~14.
\newblock Siam, 2002.

\bibitem{faria:00}
Teresa Faria.
\newblock On a planar system modelling a neuro network with memory.
\newblock {\em Journal of Differential Equations}, 168:129--149, 2000.

\bibitem{galam2012sociophysics}
Serge Galam.
\newblock {\em Sociophysics: a physicist's modeling of psycho-political
  phenomena}.
\newblock Springer, 2012.

\bibitem{giannakopoulos1999local}
Fotios Giannakopoulos and Andreas Zapp.
\newblock Local and global hopf bifurcation in a scalar delay differential
  equation.
\newblock {\em Journal of mathematical analysis and applications},
  237(2):425--450, 1999.

\bibitem{guckenheimer-holmes:83}
J.~Guckenheimer and P.~J. Holmes.
\newblock {\em Nonlinear Oscillations, Dynamical Systems and Bifurcations of
  Vector Fields}, volume~42 of {\em Applied mathematical sciences}.
\newblock Springer, 1983.

\bibitem{haidt2017disagreeing}
Jonathan Haidt.
\newblock Disagreeing virtuously.
\newblock {\em Disagreeing Virtuously}, page 138, 2017.

\bibitem{hale-lunel:93}
J.K. Hale and S.M.V. Lunel.
\newblock {\em Introduction to functional differential equations}.
\newblock Springer Verlag, 1993.

\bibitem{hermann2012heterogeneous}
Geoffroy Hermann and Jonathan Touboul.
\newblock Heterogeneous connections induce oscillations in large-scale
  networks.
\newblock {\em Physical review letters}, 109(1):018702, 2012.

\bibitem{jackson2016market}
Antony Jackson and Daniel Ladley.
\newblock Market ecologies: The effect of information on the interaction and
  profitability of technical trading strategies.
\newblock {\em International Review of Financial Analysis}, 47:270--280, 2016.

\bibitem{juul2017hipsters}
Jonas~S Juul and Mason~A Porter.
\newblock Hipsters on networks: How a small group of individuals can lead to an
  anti-establishment majority.
\newblock {\em arXiv preprint arXiv:1707.07187}, 2017.

\bibitem{kuznetsov2013elements}
Yuri~A Kuznetsov.
\newblock {\em Elements of applied bifurcation theory}, volume 112.
\newblock Springer Science \& Business Media, 2013.

\bibitem{plevin:08}
Julia Plevin.
\newblock {Who's a Hipster}.
\newblock {\em The Huffington Post}, 2008.

\bibitem{pra2018climb}
Paolo~Dai Pra, Elena Sartori, and Marco Tolotti.
\newblock Climb on the bandwagon: Consensus and periodicity in a lifetime
  utility model with strategic interactions.
\newblock {\em arXiv preprint arXiv:1804.07469}, 2018.

\bibitem{quininao2015limits}
Cristobal Qui{\~n}inao and Jonathan Touboul.
\newblock Limits and dynamics of randomly connected neuronal networks.
\newblock {\em Acta Applicandae Mathematicae}, 136(1):167--192, 2015.

\bibitem{roxin-brunel-etal:05}
A.~Roxin, N.~Brunel, and D.~Hansel.
\newblock {Role of Delays in Shaping Spatiotemporal Dynamics of Neuronal
  Activity in Large Networks}.
\newblock {\em Physical Review Letters}, 94(23):238103, 2005.

\bibitem{scheutzow1985noise}
Michael Scheutzow.
\newblock Noise can create periodic behavior and stabilize nonlinear
  diffusions.
\newblock {\em Stochastic processes and their applications}, 20(2):323--331,
  1985.

\bibitem{scheutzow1985some}
Michael Scheutzow.
\newblock Some examples of nonlinear diffusion processes having a time-periodic
  law.
\newblock {\em The Annals of Probability}, pages 379--384, 1985.

\bibitem{sherrington-kirkpatrick:75}
D.~Sherrington and S.~Kirkpatrick.
\newblock Solvable model of a spin-glass.
\newblock {\em Physical review letters}, 35(26):1792--1796 %@ 1079--7114, 1975.

\bibitem{sompolinsky1988chaos}
Haim Sompolinsky, Andrea Crisanti, and Hans-Jurgen Sommers.
\newblock Chaos in random neural networks.
\newblock {\em Physical review letters}, 61(3):259, 1988.

\bibitem{strogatz2018nonlinear}
Steven~H Strogatz.
\newblock {\em Nonlinear dynamics and chaos: with applications to physics,
  biology, chemistry, and engineering}.
\newblock CRC Press, 2018.

\bibitem{sznitman1991topics}
Alain-Sol Sznitman.
\newblock Topics in propagation of chaos.
\newblock In {\em Ecole d'Et{\'e} de Probabilit{\'e}s de Saint-Flour XIX},
  pages 165--251. Springer, 1991.

\bibitem{touboul2012limits}
Jonathan Touboul.
\newblock Limits and dynamics of stochastic neuronal networks with random
  heterogeneous delays.
\newblock {\em Journal of Statistical Physics}, 149(4):569--597, 2012.

\bibitem{touboul2014hipster}
Jonathan Touboul.
\newblock The hipster effect: When anticonformists all look the same.
\newblock {\em arXiv preprint arXiv:1410.8001}, 2014.

\bibitem{touboul2012noise}
Jonathan Touboul, Geoffroy Hermann, and Olivier Faugeras.
\newblock Noise-induced behaviors in neural mean field dynamics.
\newblock {\em SIAM Journal on Applied Dynamical Systems}, 11(1):49--81, 2012.

\end{thebibliography}

\end{document}